\documentclass[mnsc,nonblindrev,copyedit]{informs3a} 

\usepackage{preamble}

\newcommand{\bthetahat}{\hat{\boldsymbol{\theta}}}

\newcommand{\by}{\boldsymbol{y}}

\newcommand{\bmu}{\boldsymbol{\mu}}
\newcommand{\bSigma}{\boldsymbol{\Sigma}}
\newcommand{\bI}{\boldsymbol{I}}
\newcommand{\bv}{\boldsymbol{v}}

\newcommand{\cH}{\mathcal{H}}

\newcommand{\cS}{\mathcal{S}}
\newcommand{\cC}{\mathcal{C}}

\DeclareMathOperator{\tr}{tr}

\begin{document}
% \RUNAUTHOR{}
% \RUNTITLE{}
\TITLE{\Large A Solicit-Then-Suggest Model of Agentic Purchasing}

\ARTICLEAUTHORS{
  \AUTHOR{Shengyu Cao}
 \AFF{Rotman School of Management, University of Toronto, \EMAIL{shengyu.cao@rotman.utoronto.ca}}
 \AUTHOR{Ming Hu}
 \AFF{Rotman School of Management, University of Toronto, \EMAIL{ming.hu@rotman.utoronto.ca}}
}

\ABSTRACT{The e-commerce landscape is undergoing a fundamental shift from search-based shopping to agentic purchasing. AI shopping agents are replacing keyword search with multi-round conversations, in which the agent learns what customers want through targeted questions and then presents a tailored assortment of recommendations. We develop the solicit-then-suggest framework to model this emerging paradigm of agentic purchasing. An agent conducts $m$ rounds of targeted solicitation in a $d$-dimensional preference space to progressively refine its belief about the customer's ideal product, and then recommends $k$ products from which the customer selects. We identify the economic forces shaping this new form of e-commerce. Under a Gaussian prior for the base model, our analysis reveals an uncertainty decomposition identity: solicitation depth and assortment breadth are formal substitutes, and total prior uncertainty is a fixed travesty divided between the uncertainty that solicitation eliminates and the uncertainty that tailored assortment breadth hedges. Somewhat surprisingly, the two instruments improve match quality at sharply different rates: expected loss shrinks at a rate of $O(1/m)$ with solicitation depth, whereas tailored assortment breadth suffers a curse of dimensionality, with loss shrinking at the slower rate of $O(k^{-2/d})$. A few well-targeted questions can thus achieve what would otherwise require an exponential number of recommended products. The optimal assortment forms a Voronoi partition, placing each product at the posterior centroid of the candidate ideal positions it targets. For a single-product assortment, the optimal solicitation follows a water-filling rule that equalizes posterior uncertainty across preference dimensions. When multiple products are available, the coordinated optimum can exhibit selective focus, allocating less precision to directions that the assortment can hedge. Moreover, the water-filling policy designed for a single-product assortment admits a general approximation guarantee for larger assortments, with an efficiency gap that vanishes as the preference dimension grows. Beyond the Gaussian case, we show that the uncertainty decomposition identity and the substitutability between solicitation depth and assortment breadth extend to non-Gaussian priors, where the Gaussian model serves as a conservative benchmark and becomes asymptotically exact as the conversation lengthens.}
%\KEYWORDS{}
%\HISTORY{This version: \today}
\maketitle

\section{Introduction}
The e-commerce landscape is undergoing a seismic shift from product search to conversation-based agentic purchasing. A new generation of AI shopping agents engages customers through multi-round natural-language conversations rather than keyword search: a customer describes what they want, the agent asks targeted follow-up questions, and then presents a curated assortment of recommendations.\footnote{Major deployments include Amazon Rufus (February 2024), OpenAI's conversational shopping feature (April 2025), Google's agentic checkout (November 2025), and Shopify's agentic storefronts (January 2026).} The shift is already at scale: 41\% of consumers now use natural language rather than keywords when shopping online, and AI-assisted queries run 23 times longer than traditional keyword searches.\footnote{Bloomreach Consumer Survey, ``How AI is Changing Online Shopping Behavior,'' June 2025. Google VP Vidhya Srinivasan reported the query-length figure at the Google Search Central conference, November 2025.} McKinsey estimates that AI agents could mediate \$3-5 trillion of global consumer commerce by 2030.\footnote{McKinsey \& Company, ``The Agentic Commerce Opportunity,'' October 2025.} In this paper, we develop the \emph{solicit-then-suggest} framework, a formal model of agentic purchasing that captures the interplay between conversational preference learning and assortment design. The goal is to uncover the economic principles that govern this new paradigm: what forces shape how an AI agent learns customer preferences through conversation, and how does this fundamentally differ from traditional product search and recommendation?

\begin{figure}[!htb]
    \centering
    \subfigure[Traditional product search]{\includegraphics[width=0.48\linewidth]{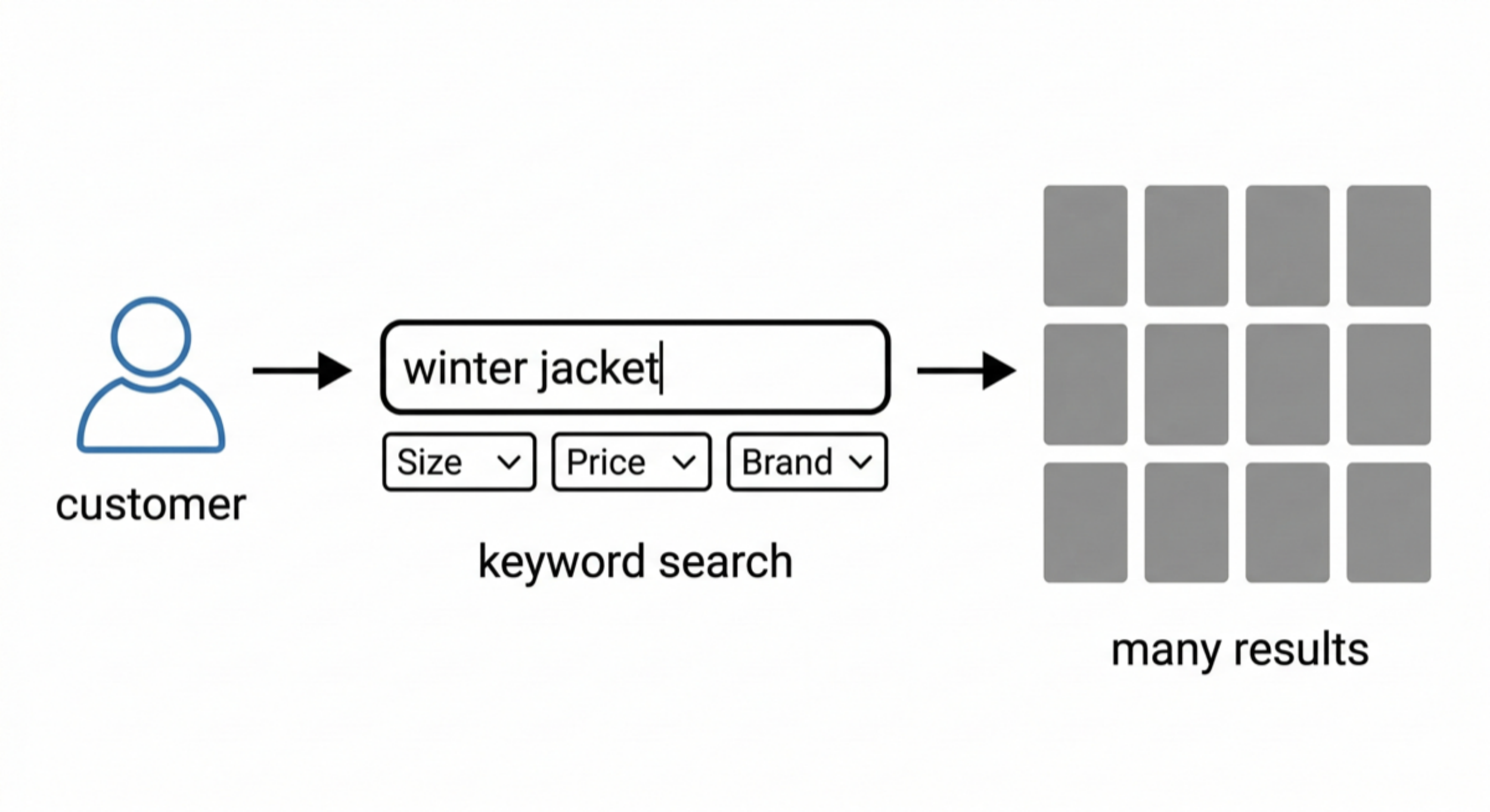}}
    \hfill
    \subfigure[Agentic purchasing (solicit-then-suggest)]{\includegraphics[width=0.48\linewidth]{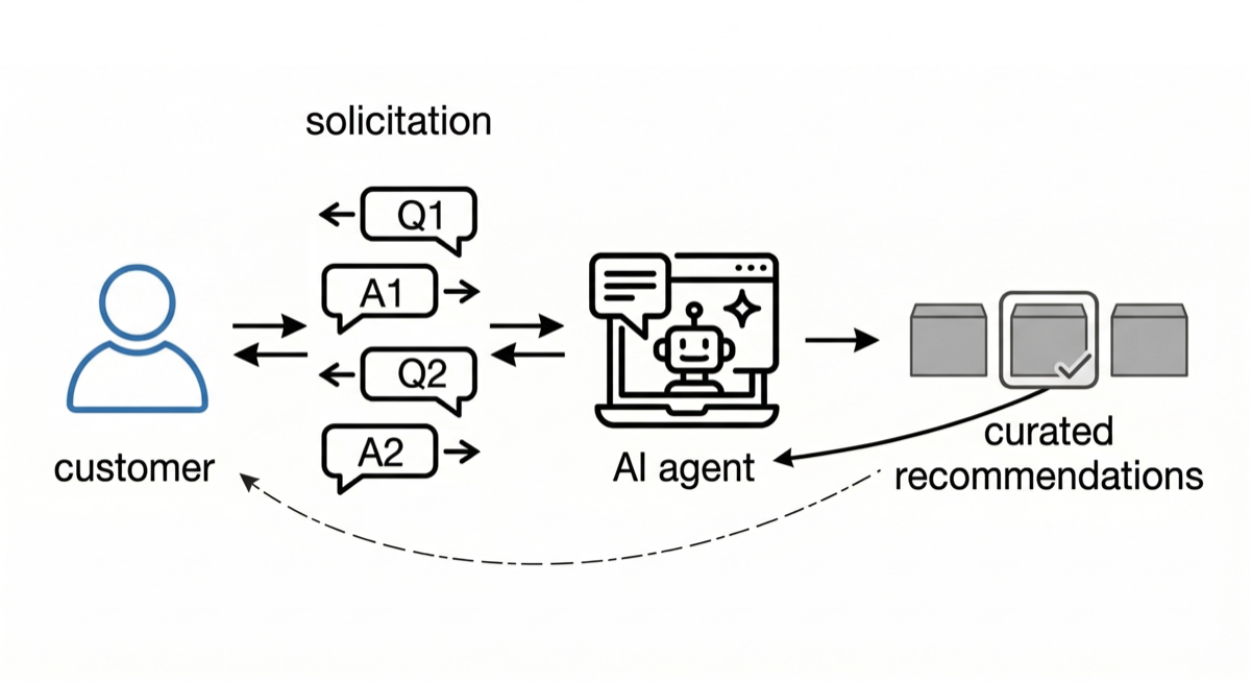}}
    \caption{Traditional product search versus agentic purchasing.\label{fig:search-vs-agentic}}
{\begin{flushleft}\footnotesize{\it Note.} In~(a), a customer enters keywords and browses a long list of results with limited ability to express nuanced preferences. In~(b), a customer converses with an AI agent across multiple rounds of targeted questioning, after which the agent presents a small, curated assortment.\end{flushleft}}
\end{figure}

Three features distinguish agentic purchasing from traditional search or structured preference surveys (Figure~\ref{fig:search-vs-agentic}). First, conversational AI extracts far richer preference information than keyword search. When interacting with an agent, a customer naturally describes style, context, use case, and trade-offs (``cozy but not bulky, warm enough for a Chicago winter but stylish for the office''), and actively corrects mistakes (``more like a puffer jacket, not wool''). Traditional e-commerce offers only a search bar and a handful of filters. Second, natural language descriptions of taste \emph{scale} in a way that rankings and comparisons do not: once a customer has articulated preferences, the marginal cost for an AI agent to evaluate one additional product is near zero, whereas asking a human to rank or compare many options quickly becomes cognitively exhausting. This asymmetry between the agent's cheap evaluation and the customer's costly selection is a fundamental economic feature that our model formalizes, and that drives several of our key results. Third, customers remain actively in the loop throughout the interactions, answering, correcting, and refining. The resulting cooperative structure is what makes multi-round solicitation effective and distinguishes this process from a one-shot recommendation engine that learns passively from past click and purchase data. As AI agents move from experimental prototypes to mainstream commerce in 2025 and 2026, theoretical foundations for their design become urgent. Early experimental evidence already confirms that agent-mediated markets differ qualitatively from human-driven commerce. \citet{AllouahEtAl2025} deploy frontier AI models as autonomous shoppers in a controlled e-commerce sandbox and find that demand concentrates on a few modal products, market shares shift dramatically across model updates, and agents respond to product positioning and platform endorsements in systematic but model-specific ways. These empirical patterns underscore the importance of understanding, at a foundational level, how the agent's preference-learning and assortment-design choices shape market outcomes.

Existing work on information acquisition, preference elicitation, and assortment optimization treats solicitation and assortment as separate problems. Traditional e-commerce relies solely on assortment to match customers with products. Agentic purchasing introduces a second instrument, active solicitation, and joint analysis of the two reveals a central insight about their relative power. Solicitation depth (asking more questions) and assortment breadth (offering more products) both improve the match between customers and products, but they are fundamentally substitutes. We establish an \emph{uncertainty decomposition identity} showing that the total prior uncertainty is a fixed travesty, divided between the uncertainty that solicitation eliminates and the uncertainty that assortment breadth hedges. The two instruments are therefore substitutes, but they operate at drastically different rates. Under optimal solicitation, the expected loss decays at a rate of $O(1/m)$, where $m$ is the number of questions. Each additional product reduces it at a rate of $O(k^{-2/d})$, where $k$ is the number of products and $d$ is the dimension of the preference space (referred to as the preference dimensionality). Assortment breadth suffers from the curse of dimensionality, whereas solicitation depth avoids it entirely. This rate asymmetry provides a formal foundation for restraint on recommendation assortment size, complementing behavioral evidence on choice overload \citep{IyengarLepper2000} and echoing the cheap-AI-evaluation-costly-selection asymmetry described above. A few well-targeted questions are worth more than a long list of options.

% This paper develops a model of AI-assisted purchasing, the \emph{solicit-then-suggest} model, that captures the simultaneous design of solicitation and recommendation. The remainder of the paper is organized as follows. Section~\ref{sec:literature} reviews related literature. Section~\ref{sec:model} presents the model. Section~\ref{sec:analysis} develops the main analysis under a Gaussian prior, covering optimal recommendation (Section~\ref{sec:recommendation}), optimal solicitation policy (Section~\ref{sec:solicitation}), and the interplay between solicitation and recommendation (Section~\ref{sec:interplay}). Section~\ref{sec:non-gaussian} extends the results to general priors and provides a geometric interpretation through partition trees. Section~\ref{sec:conclusion} offers concluding remarks and directions for future work.

{\bf Model.}
An AI agent assists a customer in finding a product in a $d$-dimensional feature space. The customer has a latent ideal point representing the customer's most preferred product, and utility decreases in the squared Euclidean distance between the recommended product and this ideal point \citep{Hotelling1929, Lancaster1966}. The agent begins with a prior belief about the ideal point, formed from processing the customer's initial natural-language description. We assume a Gaussian prior for the base model and extend key results to general priors. The agent then conducts $m$ rounds of targeted questioning: in each round, the agent selects a direction in the feature space and observes a noisy signal of the customer's ideal position along that direction. After all $m$ rounds, the agent recommends an assortment of $k$ products, and the customer selects the one closest to their ideal point. The agent's problem is to choose the $m$ query directions (the solicitation policy) and the $k$ products (the assortment rule) to maximize expected customer utility.

{\bf Results.}
We begin with the assortment problem, taking the agent's posterior belief as given. Under any prior with finite second moments, when the agent recommends a single product, the optimal choice is the posterior mean, and the expected loss equals half the total posterior variance. This clean separation between bias and variance transforms the agent's questioning problem into a pure variance-reduction objective.

Under the Gaussian benchmark, when the agent recommends multiple products, the optimal assortment partitions the preference space into regions and places each product at the centroid of its region. In the two-product case, the optimal pair is placed symmetrically along the direction of greatest remaining posterior uncertainty, and the improvement over a single product is proportional to this remaining uncertainty. For a practitioner, this means the assortment should hedge against whatever the solicitation has left unresolved. With more products, the assortment can hedge across additional directions of uncertainty, but the expected loss shrinks only at a rate $O(k^{-2/d})$.

We then turn to the solicitation problem. For a single-product assortment, the agent's objective reduces to minimizing the agent's total remaining uncertainty about the customer's ideal point. The optimal policy follows a water-filling rule: the agent targets the preference dimensions of greatest prior uncertainty and allocates questions until the remaining uncertainty is equalized across all actively learned dimensions. When the agent can recommend multiple products, the single-product water-filling policy remains a useful benchmark. Under equal prior uncertainty across all preference dimensions, this policy is exactly optimal for two products when fewer questions are available than the preference dimensionality, and its efficiency loss vanishes as the preference dimensionality grows. More generally, we provide an approximation guarantee for using the $k=1$ policy with larger assortments. This near-optimality result has immediate practical value, as it means that system designers can often use a simple solicitation policy without solving the full joint problem from scratch.

When the two instruments are jointly designed, the substitutability between solicitation depth and assortment breadth holds for general $k$: each additional question reduces the full-information gap, thereby shrinking the scope for assortment breadth to add value. Because solicitation depth avoids the curse of dimensionality that governs assortment breadth, moderate questioning renders large assortments unnecessary. In the two-product case, when sufficient rounds of solicitation are available, the joint optimum can exhibit \emph{selective focus}: the optimal solicitation asks fewer questions along the direction in which the assortment can hedge and concentrates on the remaining dimensions. This yields a concrete managerial prescription: invest in targeted questioning to learn preferences, then present a small, curated set of options.

Beyond the Gaussian case, we extend the value decomposition and uncertainty decomposition identity to general priors: under any prior with finite variance, the substitutability between solicitation depth and assortment breadth continues to hold. When the agent uses a predetermined sequence of questions, the Gaussian model provides a conservative benchmark that bounds the solicitation value from below. As the conversation lengthens and explores all preference dimensions, the posterior converges to a Gaussian regardless of the initial prior, and the closed-form Gaussian formulas become asymptotically exact.

\section{Literature Review}\label{sec:literature}

Our work is closely related to four streams of literature: AI-assisted purchasing and preference learning, information acquisition and experimental design, feature-based pricing and market segmentation, and assortment optimization.

{\bf AI-assisted purchasing and preference learning.}
A growing body of literature examines how AI agents learn customer preferences and act on these learnings. \citet{DongJhunjhunwalaKanoria2025} model the interaction between a customer and an AI recommender through a rational inattention lens: the customer sends a single costly message about preferences, and the agent forms a posterior belief and draws recommendations from it, jointly optimizing message precision and the size of the recommendation set. Their key insight is that, in high dimensions, communication and search must be optimized jointly, as neither alone suffices. On the empirical side, \citet{AllouahEtAl2025} build a controlled sandbox for auditing how AI shopping agents select products and document that current agents exhibit choice concentration, demand patterns that shift sharply across AI models, and strong sensitivity to product positioning. These findings provide early evidence of the distinctive purchasing patterns exhibited by AI agents. On the preference estimation side, \citet{AouadElGadarriFarias2025} show that standard reward-learning methods for AI alignment are vulnerable to preference heterogeneity: pooling pairwise comparison data from diverse users and fitting a single utility model systematically overweighs uncertain users and discounts those with strong preferences. Their sign estimator, a simple modification to the loss function, corrects this bias and recovers population-average ordinal preferences. On the modeling side, our solicitation model builds on the sequential questioning structure pioneered by conjoint analysis in marketing science. The conjoint tradition has progressively shifted from static survey designs \citep{Green1990} to polyhedral methods that tailor each question to previous answers \citep{ToubiaSimesterHauserDahan2003,ToubiaHauserSimester2004}, and, most recently, to Bayesian adaptive frameworks that use Gaussian posterior approximations and account for response noise \citep{SaureVielma2019}. Our model departs from the above works. Relative to \citet{AouadElGadarriFarias2025}, who study how to aggregate preferences across a heterogeneous population, our agent operates at the individual level, taking its prior as given and refining it for a single customer through solicitation and assortment design. Relative to the conjoint tradition, which optimizes questions to estimate the full preference vector, our agent jointly optimizes solicitation and a specific downstream assortment decision. Relative to the AI-assisted search model of \citet{DongJhunjhunwalaKanoria2025}, which features one-shot communication, our solicit-then-suggest model captures multi-round adaptive conversation and establishes that solicitation depth and assortment breadth are formal substitutes, linked by an uncertainty decomposition identity. This substitutability carries a pronounced rate asymmetry: expected mismatch decays at rate $O(1/m)$ in solicitation depth but only at rate $O(k^{-2/d})$ in assortment breadth, so that solicitation avoids the curse of dimensionality that governs assortment expansion.

{\bf Information acquisition and experimental design.}
The information acquisition literature studies how decision-makers allocate attention across information sources under processing constraints. \citet{Sims2003} introduces the rational inattention framework, in which agents face a finite information-processing capacity and must choose which signals to attend to. \citet{VanNieuwerburghVeldkamp2010} apply this framework to portfolio choice and show that capacity-constrained investors optimally specialize their information acquisition, leading to under-diversification. \citet{CheMierendorff2019} characterize optimal dynamic attention allocation across biased news sources prior to a binary decision and show that rich dynamics arise from the interaction between beliefs and attention costs. \citet{LiangMuSyrgkanis2022} study an agent who dynamically allocates attention across multiple information sources, each informing a different component of an unknown multi-dimensional Gaussian state, and characterize the optimal acquisition strategy for a downstream binary choice. \citet{Zhong2022} develops a unified framework in which a decision-maker freely designs any dynamic signal process before choosing among alternatives, and shows that the optimal policy has a Poisson-bandits structure. A complementary strand of the experimental design literature develops criteria for selecting observations to maximize statistical efficiency, with A-optimality (trace minimization) and D-optimality (determinant minimization) as leading criteria \citep{Pukelsheim2006}, recently unified through an optimization lens by \citet{Zhao2024}. Within sequential experimental design, \citet{Zhao2026} studies how to adaptively allocate experimental units across treatment arms to minimize variance by using earlier observations to guide later allocation decisions, and \citet{ZhaoZhou2025} propose a partition-based design that divides the covariate space into cells and balances treatment assignment within each cell through an online matching formulation. The sequential allocation of a limited budget across dimensions in these designs parallels our agent's solicitation problem, and the partition-based balancing in \citet{ZhaoZhou2025} mirrors the Voronoi partition that underlies our recommendation rule. Our solicitation problem inherits this sequential structure: the agent allocates a finite budget of queries across preference dimensions. For a single recommendation ($k=1$), the problem reduces to the A-optimal design, but the trace-minimization criterion is not a modeling choice as in classical experimental design. It emerges endogenously from the customer's utility because minimizing expected mismatch is equivalent to minimizing the posterior trace. Our formulation differs from classical approximate designs in that each question is a sequential rank-one precision update rather than a continuous allocation, and we show that a rank-capped water-filling policy is the exact optimal solution. For multiple recommendations ($k\ge 2$), the downstream assortment fundamentally changes the solicitation problem. Classical experimental design and the information acquisition papers above either target pure estimation accuracy with no downstream action or optimize data collection for a single selection from a given set of alternatives. Our agent, by contrast, jointly designs the information policy and a multi-product assortment that it will offer. This anticipation reshapes the solicitation policy: the agent can deliberately maintain higher uncertainty along directions the assortment can hedge, and reallocate effort to dimensions that only questioning can resolve, a phenomenon we call selective focus and demonstrate explicitly in the two-product case. The resulting interaction produces formal substitutability between solicitation depth and assortment breadth, with a rate asymmetry that favors solicitation: expected mismatch after $m$ questions decays at rate $O(1/m)$, while the gain from $k$ products decays at rate $O(k^{-2/d})$, so a few well-targeted questions outperform exponentially many products.

{\bf Feature-based pricing and market segmentation.}
A related body of literature studies how firms use product or customer features to learn valuations and partition markets. \citet{CohenLobelPaesLeme2020} consider a firm that prices differentiated products online, each described by a feature vector, when the mapping from features to market values is unknown. Each sale or no-sale outcome narrows what the firm knows about feature values, and, geometrically, each such outcome corresponds to a hyperplane cutting the set of plausible parameter values, with the cut direction determined by the arriving product's features. By maintaining an ellipsoid that tracks all parameter values consistent with past observations, the firm learns feature values quickly, achieving regret that scales as $O(d^2 \log T)$, only logarithmic in the selling horizon $T$. \citet{CuiHamilton2025} study how a seller who has already learned a predictive model of customer valuations should partition the customer population into $k$ market segments, each offered a tailored price. They show that the revenue loss from using $k$ segments instead of fully personalized pricing is at most $O(1/k)$, and that revenue is concave in the number of segments, so the gains from finer segmentation diminish. These papers connect to our model through a shared geometric structure. The accept-reject signal in \citet{CohenLobelPaesLeme2020} plays the same role as a solicitation query in our model. Both cut through a region of uncertainty about preferences with a hyperplane, and each subsequent observation shrinks the plausible set. The key difference is who controls the cut direction. In feature-based pricing, the direction is determined by the arriving product, whereas our agent can actively choose the direction of each query. On the segmentation side, the $k$-segment partition in \citet{CuiHamilton2025} parallels our $k$-product recommendation set, which partitions the posterior belief into Voronoi cells. In both settings, coarser partitions lose information at a rate governed by the number of cells. Our contribution relative to this literature is to connect these two elements: by giving the agent control over the cut directions and coupling learning with the downstream partition, we show that the quality of the partition reshapes which cuts the agent should make, and vice versa.

{\bf Assortment optimization.}
The assortment optimization literature in operations management studies how a firm should select a product set to serve a heterogeneous customer population \citep{KokFisherVaidyanathan2009}. \citet{GolrezaeiNazerzadehRusmevichientong2014} demonstrate that personalizing assortments based on observable customer characteristics can substantially improve revenue, and develop index-based policies with competitive-ratio guarantees. \citet{BernsteinModaresiSaure2019} propose a dynamic clustering approach that adaptively segments customers from transaction data to expedite preference learning for assortment personalization. \citet{AgrawalAvadhanula2019} formulate the exploration-exploitation tradeoff in dynamic assortment selection as a bandit problem under the multinomial logit choice model, and derive regret bounds for UCB-based algorithms. \citet{LongEtAl2025} conduct a field experiment with 1.6 million customers, and show that purchase probability first increases and then decreases with assortment size. This provides direct evidence for choice overload in online recommendation systems. Our assortment problem shares the partition-and-target logic of classical assortment planning: the agent's posterior belief serves as a type distribution, and the optimal assortment partitions the type space into Voronoi cells, with each product at the centroid of its cell. We show that this optimization is equivalent to vector quantization \citep{GershoGray1992}. Two departures from classical assortment are central. First, the type distribution is endogenous: the agent shapes it through solicitation rather than taking customer heterogeneity as given. Second, solicitation and assortment breadth are formal substitutes, linked by an uncertainty decomposition identity that splits total prior uncertainty between what solicitation eliminates and what assortment can hedge. This substitutability carries a pronounced rate asymmetry: expected mismatch after $m$ questions decays at rate $O(1/m)$, while the gain from $k$ products decays at rate $O(k^{-2/d})$, so a few well-targeted questions achieve what would require exponentially many products in high dimensions. This provides a formal foundation for small, curated assortments when the agent can learn preferences through conversation.

\section{Model}\label{sec:model}
\begin{figure}[htp]
\centering
\includegraphics[width=0.9\textwidth]{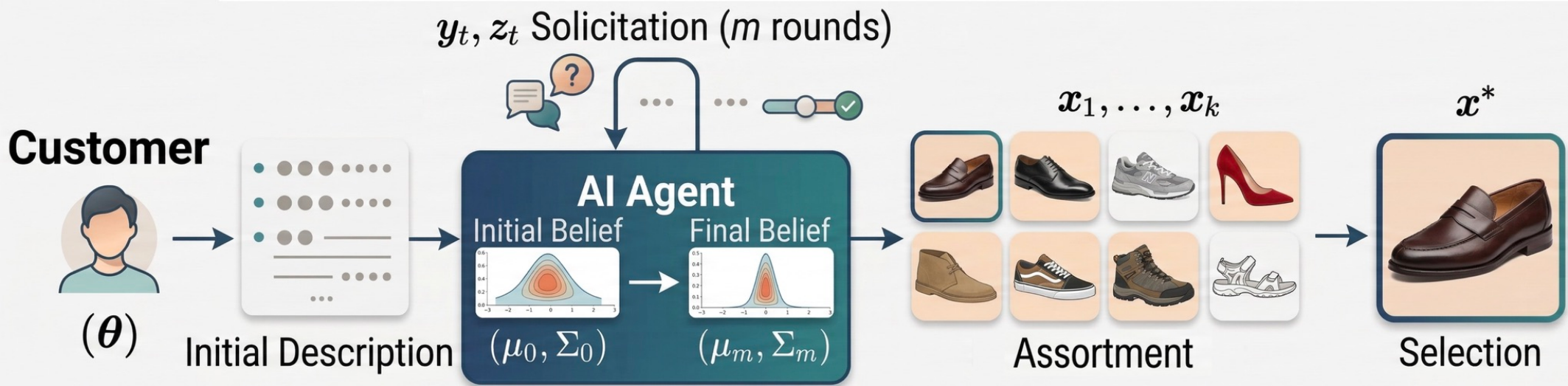}
\caption{The solicit-then-suggest model.} 
\label{fig:process-flow}
{\begin{flushleft}\footnotesize{\it Note.} The customer provides an initial description, the agent conducts $m$ rounds of preference solicitation (exchanging query directions $\by_t$ and noisy responses $z_t$), and then recommends an assortment of $k$ products from which the customer selects.\end{flushleft}}
\end{figure}

We consider an AI shopping agent that assists a customer in finding a suitable product through a sequence of interactions. The customer provides an initial description, the agent conducts $m$ rounds of preference solicitation, and then recommends an assortment of $k$ products. We call this the \emph{solicit-then-suggest} model. Each product is represented by a feature vector $\bx \in \mathbb{R}^d$, and we treat the product space as continuous, allowing the agent to recommend any point in $\mathbb{R}^d$. This is a tractability assumption appropriate when the product catalog is large and densely covers the relevant feature region. In modern e-commerce, where major platforms list hundreds of millions of products\footnote{Amazon reportedly lists over 350 million products as of 2024 (Marketplace Pulse, ``Amazon Product and Seller Count,'' February 2024).} and mass customization enables near-arbitrary attribute combinations, any theoretically optimal point can be closely approximated by an available product. Figure~\ref{fig:process-flow} summarizes the interaction.

{\bf Customer preferences and utility.}
Each customer is characterized by an ideal product vector $\btheta \in \mathbb{R}^d$, which represents their most preferred point in the feature space. This specification aligns with the ideal-point model in marketing science \citep{Hotelling1929} and is widely used in choice theory. The customer obtains utility from product $\bx$ according to
$$
U(\btheta, \bx) = -\frac{1}{2}\|\bx - \btheta\|_2^2.
$$
If $\btheta$ were known, the optimal recommendation $\bx^* = \btheta$ would achieve the optimal utility of zero. However, since the agent does not directly observe $\btheta$, it acquires information through the following two sources.

{\bf Initial description.}
At the outset, the customer provides a natural language description of their needs (e.g., ``I'm looking for a lightweight laptop for travel''). In practice, this initial prompt may be supplemented by a small number of structured follow-up questions, such as a budget range or an intended use case, that the customer can answer or skip. The agent may also draw on persistent memory retained from prior interactions, such as past purchases and previously expressed preferences. Cross-session memory of this kind is increasingly offered by major AI assistants. A returning customer with rich interaction history provides the agent with more precise prior knowledge, while a first-time customer with no stored memory induces greater initial uncertainty. We refer to the customer's initial prompt, any answered follow-up questions, and any stored memory collectively as the \emph{initial description}. The agent processes this description to form an initial estimate $\bmu_0 \in \mathbb{R}^d$ of the ideal point, with associated uncertainty captured by a covariance matrix $\bSigma_0 \in \mathbb{R}^{d \times d}$. We assume $\bSigma_0 \succ 0$ without loss of generality, as dimensions with zero prior variance are already known and can be excluded beforehand. Accordingly, throughout the paper, $d$ denotes the effective dimension that remains uncertain after this initial-description stage.

{\bf Sequential preference solicitation.}
After processing the initial description, the agent enters a solicitation phase with $m$ additional rounds to narrow down the customer's ideal point $\btheta$. In each round $t \in \{1, \ldots, m\}$:
\begin{enumerate}
    \item The agent selects a query direction $\by_t \in \mathbb{R}^d$ with $\|\by_t\|_2 = 1$.
    \item The customer provides a noisy response revealing their ideal point along this direction:
    $$
        z_t = \btheta^\top \by_t + \epsilon_t,
    $$
    where $\epsilon_t \sim \mathcal{N}(0, \sigma^2)$ are i.i.d.\ noise terms.
\end{enumerate}
Intuitively, the query $\by_t$ encodes a question about the customer's preferences: its entries specify the weight assigned to each feature dimension. The signal $\btheta^\top \by_t = \sum_i y_{t,i}\,\theta_i$ aggregates the customer's ideal feature levels according to these weights. When $\by_t$ equals a standard basis vector $\boldsymbol{e}_i$ (e.g., the screen-size dimension for a laptop), the question targets a single feature, and $z_t = \theta_i + \epsilon_t$ directly reveals the customer's ideal level of that feature. When $\by_t$ combines features with opposing signs (e.g., portability versus performance), the response reveals the customer's preferred tradeoff. The unit-norm constraint $\|\by_t\|_2 = 1$ normalizes each question so that observation noise variance is $\sigma^2$ regardless of which features are queried, reducing the agent's problem to choosing which combination of features to ask about. The noise term $\epsilon_t$ captures the customer's uncertainty about their preferences, ambiguity in natural-language communication, and interpretation errors in the language model that translates the response into a numerical score. We denote by $\cH_t = (\bmu_0, \bSigma_0, \by_1, z_1, \ldots, \by_t, z_t)$ the information available after round $t$, with $\cH_0 = (\bmu_0, \bSigma_0)$.

\begin{figure}[htp]
\centering
\subfigure[Preview-and-rate card. The agent presents a candidate product and solicits a binary response (``Not interested'' vs. ``More like this''), with an option to skip.]{\includegraphics[width=0.495\textwidth]{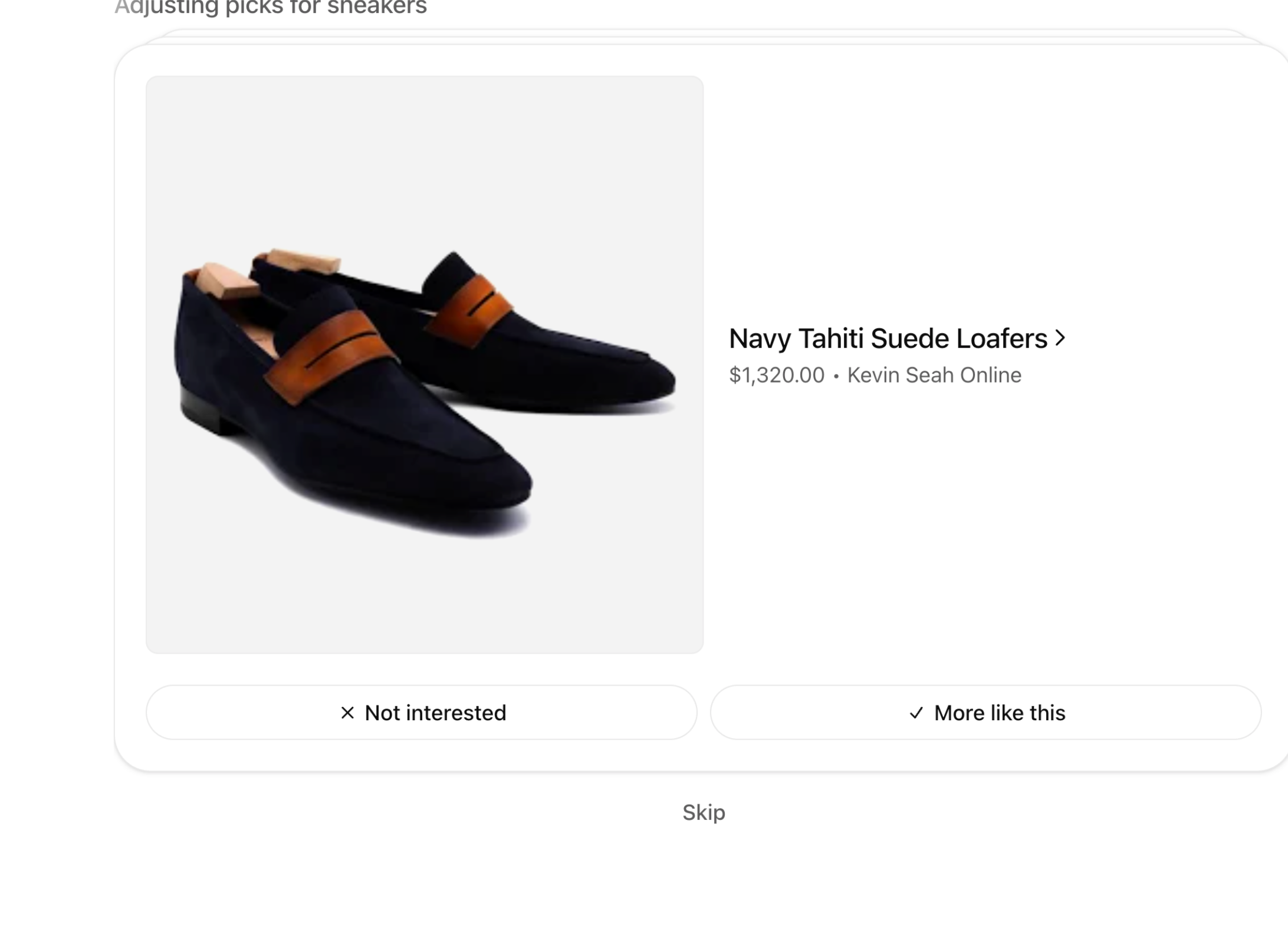}}
\hfill
\subfigure[Illustrative follow-up. Following a dislike, the interface may ask which attribute (price, style, brand, or features) drove the response.]{\includegraphics[width=0.495\textwidth]{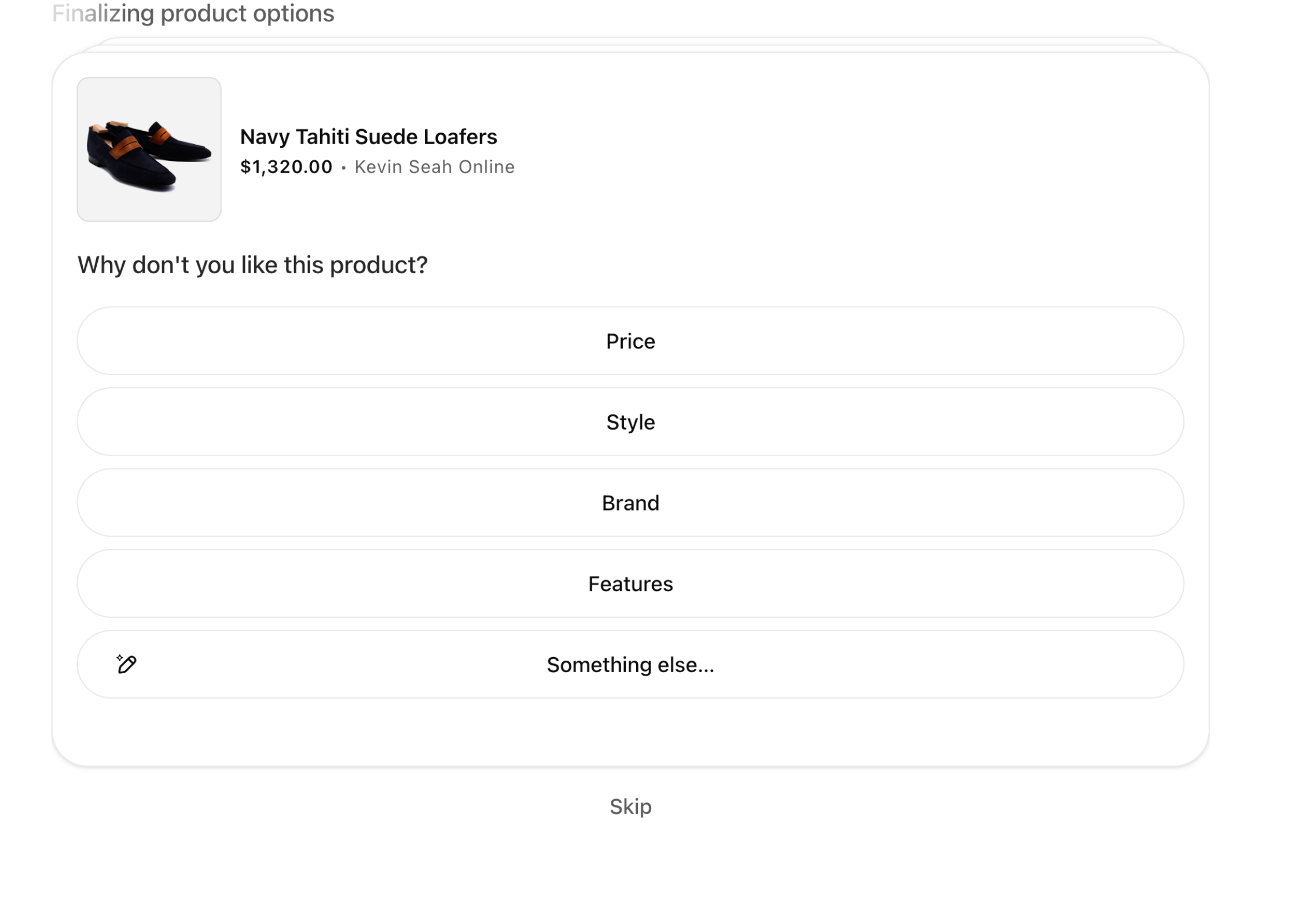}}
\caption{Solicitation interface from OpenAI's shopping agent in ChatGPT (screenshots captured by the authors, January 2026).}
{\begin{flushleft}\footnotesize{\it Note.} The model abstracts from this specific UI and instead treats each round as one normalized opportunity to extract information before the assortment stage.\end{flushleft}}
\label{fig:ui-solicitation}
\end{figure}

This sequential solicitation model generalizes the adaptive conjoint analysis framework \citep{Green1990} to a setting in which an AI agent learns preferences via targeted queries and uses these learnings to design recommendations. In practice, agents may implement a solicitation round by presenting a candidate product or a targeted prompt, observing coarse responses such as ``more like this'' or ``not interested,'' and using a brief clarification to interpret the reaction. Figure~\ref{fig:ui-solicitation} illustrates one such implementation from OpenAI's shopping agent in ChatGPT. The unit-norm query $\by_t$ and scalar response $z_t$ together form a reduced-form representation of the net information content of one interaction round, a standard approach in information acquisition models \citep{Sims2003, VanNieuwerburghVeldkamp2010}.

{\bf Agent's decision problem.}
After the $m$ solicitation rounds, the agent recommends an assortment of $k$ products $\cS \subset \mathbb{R}^d$. The customer selects the product from this assortment that is closest to their ideal point, yielding utility
$$
    V(\cS; \btheta) = \max_{\bx \in \cS} U(\btheta, \bx) = -\frac{1}{2}\min_{\bx \in \cS} \|\bx - \btheta\|_2^2.
$$
Since better-matched recommendations directly benefit the customer, we model the customer as a cooperative respondent who answers queries truthfully, though possibly with imprecise or inaccurate interpretation by the agent, and selects the utility-maximizing product from the assortment using their ideal point.

Offering multiple products ($k > 1$) allows the agent to hedge against estimation error in $\btheta$. The agent makes two interrelated decisions:
\begin{enumerate}
    \item \textbf{Solicitation Policy:} A sequence of mappings $\pi = (\pi_1, \ldots, \pi_m)$, where $\pi_t: \cH_{t-1} \to \{\by \in \mathbb{R}^d : \|\by\|_2 = 1\}$ specifies the query direction for round $t$.
    \item \textbf{Assortment Rule:} A mapping $\rho: \cH_m \to (\mathbb{R}^d)^k$ that determines the final assortment.
\end{enumerate}
The agent's objective is to maximize expected customer utility $\E[V(\rho(\cH_m); \btheta)]$, where the expectation is over the prior distribution of $\btheta$ and the observation noise $(\epsilon_1, \ldots, \epsilon_m)$.

\section{Analysis}\label{sec:analysis}
In the base model, we model the agent's initial belief about the customer's ideal point as a Gaussian distribution: $\btheta \sim \mathcal{N}(\bmu_0, \bSigma_0)$. The Gaussian is the maximum-entropy distribution for a given mean and covariance: among all distributions consistent with these two moments, it imposes the fewest additional assumptions. Any other distribution matching $\bmu_0$ and $\bSigma_0$ would implicitly build in extra structure, such as skewness or multimodality, that the initial description does not warrant. We extend the analysis to general priors in Section~\ref{sec:non-gaussian}.

Before proceeding, we establish the mechanism of belief updating that underpins the subsequent analysis. 
Specifically, as the observation model is linear with Gaussian noise, the posterior belief $\btheta | \cH_t$ remains Gaussian after each solicitation round. The agent's belief is therefore fully characterized at each round by two quantities: the posterior mean $\bmu_t$, which represents the best estimate of the customer's ideal point, and the posterior covariance $\bSigma_t$, which captures the residual uncertainty across feature dimensions.

\begin{lemma}[{\sc Gaussian Posterior Update}]\label{lem:posterior}
Given Gaussian prior $\btheta | \cH_{t-1} \sim \mathcal{N}(\bmu_{t-1}, \bSigma_{t-1})$ and observation $z_t = \btheta^\top \by_t + \epsilon_t$ with $\epsilon_t \sim \mathcal{N}(0, \sigma^2)$, the posterior distribution is Gaussian:
$$
\btheta | \cH_t \sim \mathcal{N}(\bmu_t, \bSigma_t),
$$
where the mean and covariance are updated via the Kalman filter equations:
\begin{align}
    \boldsymbol{\kappa}_t &= \frac{\bSigma_{t-1} \by_t}{\sigma^2 + \by_t^\top \bSigma_{t-1} \by_t}, \label{eq:kalman-gain}\\
    \bmu_t &= \bmu_{t-1} + \boldsymbol{\kappa}_t (z_t - \by_t^\top \bmu_{t-1}), \label{eq:mean-update}\\
    \bSigma_t &= \bSigma_{t-1} - \boldsymbol{\kappa}_t \by_t^\top \bSigma_{t-1}. \label{eq:cov-update}
\end{align}
Here, $\boldsymbol{\kappa}_t \in \mathbb{R}^d$ is the Kalman gain vector.
\end{lemma}

The Kalman filter provides an intuitive updating rule. The agent revises its belief by incorporating the ``surprise'' in each observation: the difference between the observed response $z_t$ and the predicted response $\by_t^\top \bmu_{t-1}$, as shown in \eqref{eq:mean-update}. The Kalman gain $\boldsymbol{\kappa}_t$ in \eqref{eq:kalman-gain} determines how much weight to place on this new information. When prior uncertainty is high (large $\bSigma_{t-1}$) or observation noise is low (small $\sigma^2$), the gain is larger, and the agent updates more significantly. Conversely, when the agent is already confident in its estimate, new observations have less influence.

Each solicitation round reduces the agent's uncertainty. By \eqref{eq:cov-update}, the update subtracts a positive semidefinite matrix from $\bSigma_{t-1}$, which ensures $\bSigma_t \preceq \bSigma_{t-1}$ for all $t \geq 1$. Moreover, when $\bSigma_{t-1}\succ 0$ and $\by_t \neq \boldsymbol{0}$, the correction term is nonzero, so $\bSigma_t \neq \bSigma_{t-1}$ and $\tr(\bSigma_t) < \tr(\bSigma_{t-1})$. This monotonic variance reduction guarantees that no solicitation round is wasted: the agent always learns from customer responses.
Following the logic of backward induction, we
first characterize the terminal-stage assortment problem, then work backward to determine
the optimal solicitation policy.

\subsection{Optimal Assortment}\label{sec:recommendation}

After $m$ solicitation rounds, the agent holds posterior belief $\btheta | \cH_m \sim \mathcal{N}(\bmu_m, \bSigma_m)$ and must select $k$ products to recommend. The customer selects the product from the assortment that maximizes their utility, yielding expected payoff $V(\cS; \btheta) = \max_{\bx \in \cS} U(\btheta, \bx)$. We define the optimal expected utility with $k$ recommendations, conditional on the solicitation history, as
$$
V_k^*(\cH_m) \;\equiv\; \max_{\cS \subseteq \R^d,\; |\cS|=k}\; \E\!\left[\left.\max_{\bx \in \cS} U(\btheta, \bx) \;\right|\; \cH_m\right].
$$
We write $V_k^*$ for brevity when the conditioning on $\cH_m$ is understood in the context. We analyze the structure of optimal assortments, beginning with the foundational single-product case before developing the general multi-product problem.

\begin{proposition}[{\sc Optimal Single-Product Assortment}]\label{prop:single-rec}
For $k=1$, the optimal recommendation is the posterior mean $\bx^* = \bmu_m$. This is because the expected loss admits the decomposition 
\begin{equation}\label{eq:bias-variance}
    \E\left[\left.\frac{1}{2}\|\bx - \btheta\|_2^2 \,\right|\, \cH_m\right] = \underbrace{\frac{1}{2}\|\bx - \bmu_m\|_2^2}_{\text{bias}} + \underbrace{\frac{1}{2}\tr(\bSigma_m)}_{\text{variance}},
\end{equation}
for any recommendation $\bx \in \mathbb{R}^d$. Setting $\bx = \bmu_m$ eliminates the bias term, yielding optimal expected utility
$$
    V_1^*(\cH_m) = -\frac{1}{2}\tr(\bSigma_m).
$$
In particular, $V_1^*(\cH_m)$ depends on the posterior covariance $\bSigma_m$ alone and is invariant to the posterior mean $\bmu_m$.
\end{proposition}

The decomposition \eqref{eq:bias-variance} separates the expected loss into two sources. The bias term $\frac{1}{2}\|\bx - \bmu_m\|_2^2$ penalizes any systematic deviation of the recommended product from the agent's best estimate, and is eliminated by setting $\bx = \bmu_m$. The variance term $\frac{1}{2}\tr(\bSigma_m)$ captures the irreducible loss from residual uncertainty about $\btheta$, and sums the posterior variances across all $d$ feature dimensions. Because these two sources are additively separable, the optimal recommendation depends only on the posterior mean (which eliminates bias), while the achievable utility depends only on the posterior covariance (which determines the residual loss).

The invariance of $V_1^*$ with respect to $\bmu_m$ has a far-reaching consequence for the solicitation phase. Since the terminal utility depends exclusively on $\bSigma_m$, the agent's $m$-round solicitation policy should be chosen to minimize $\tr(\bSigma_m)$, regardless of how the posterior mean evolves during solicitation. This transforms the dynamic sequential decision problem into a pure variance-reduction objective, where the agent seeks to shrink the total posterior variance as rapidly as possible. We formalize this connection in Section~\ref{sec:solicitation}, where we show that trace minimization coincides with the classical A-optimality criterion in experimental design and admits a clean closed-form characterization.

When the agent can recommend multiple products ($k \geq 2$), a qualitatively different mechanism becomes available: hedging against estimation error. Rather than concentrating the assortment near the posterior mean, the agent can spread products across the feature space so that, for a wider range of possible customer preferences, at least one product lies close to the customer's ideal point.
The following proposition formalizes the essential structure of optimal hedging through the two-product case, which admits a closed-form solution.

\begin{proposition}[{\sc Optimal Two-Product Hedging}]\label{prop:k2-hedging}
Fix a history $\cH_m$ with posterior $\btheta \mid \cH_m \sim \mathcal{N}(\bmu_m, \bSigma_m)$, and let $\lambda_1 \geq \cdots \geq \lambda_d > 0$ denote the eigenvalues of $\bSigma_m$ with corresponding eigenvectors $\bv_1, \ldots, \bv_d$. For $k = 2$:
\begin{enumerate}
  \item[(i)] \textbf{Hedging decomposition.} Any optimal assortment pair takes the form $\cS = \{\bmu_m - c\bv,\, \bmu_m + c\bv\}$ for some unit vector $\bv \in \mathbb{R}^d$ (the hedging direction) and scalar $c > 0$ (the spread). Let $\eta_\parallel = \bv^\top (\btheta - \bmu_m)$ denote the customer's posterior deviation along $\bv$, and let $\tau^2 = \bv^\top \bSigma_m \bv$ be the posterior variance in that direction, so that $\eta_\parallel \sim \mathcal{N}(0, \tau^2)$. The hyperplane through $\bmu_m$ orthogonal to $\bv$ partitions the feature space into two Voronoi regions $\cC_1$ and $\cC_2$, and the expected loss decomposes as
    \begin{equation}\label{eq:k2-loss}
        \E\left[\min_{\bx \in \cS} \|\bx - \btheta\|_2^2 \,\middle|\, \cH_m\right] = \underbrace{\E\left[(c - |\eta_\parallel|)^2 \,\middle|\, \cH_m\right]}_{\text{hedging residual}} + \underbrace{\tr(\bSigma_m) - \tau^2}_{\text{unhedged loss}}.
    \end{equation}
    \item[(ii)] \textbf{Optimal spread.} For a fixed hedging direction $\bv$, the hedging residual is minimized at $c^* = \E\left[|\eta_\parallel|\right] = \tau\sqrt{2/\pi}$, the expected magnitude of the customer's deviation along $\bv$. The posterior centroids of the two Voronoi regions are $\bmu_m \pm \frac{\sqrt{2/\pi}}{\tau}\,\bSigma_m\bv$, so the pair $\{\bmu_m \pm c^*\bv\}$ is centroidal if and only if $\bv$ is an eigenvector of $\bSigma_m$.
  \item[(iii)] \textbf{Optimal hedging direction and gain.} The total expected loss is minimized when $\bv = \bv_1$, the direction of greatest posterior uncertainty. The optimal assortment pair is $\cS_2^* = \bigl\{\bmu_m \pm \sqrt{2\lambda_1/\pi}\,\bv_1\bigr\}$, and this pair coincides with the posterior centroids of its Voronoi regions. Its expected utility and hedging gain are
    \begin{align}
        V_2^* &= -\frac{1}{2}\left[\left(1 - \frac{2}{\pi}\right)\lambda_1 + \sum_{i=2}^d \lambda_i\right], &
        V_2^* - V_1^* &= \frac{\lambda_1}{\pi} \approx 0.318\,\lambda_1. \label{eq:k2-gain}
    \end{align}
\end{enumerate}
\end{proposition}

The decomposition \eqref{eq:k2-loss} reveals both the power and the limitation of a second product. Two products can differ only along a single direction in feature space. Along that direction, the agent places one product on each side of its best estimate, and the customer picks the one closer to their ideal point. Because the chosen product already sits on the correct side of the agent's best estimate, the remaining mismatch along that direction, $\bigl||\eta_\parallel| - c\bigr|$, is smaller than it would be with a single recommendation. The agent controls the extent to which the mismatch is absorbed by adjusting the spread $c$ between the two products. In every other direction, however, both products are positioned identically at the agent's best estimate $\bmu_m$. Regardless of which product the customer selects, the mismatch in the remaining $d-1$ dimensions is the same. A second product, therefore, can hedge against uncertainty in one dimension of the feature space but leaves the customer fully exposed to uncertainty in the remaining dimensions.

\begin{figure}[htp]
\centering
\includegraphics[width=\textwidth]{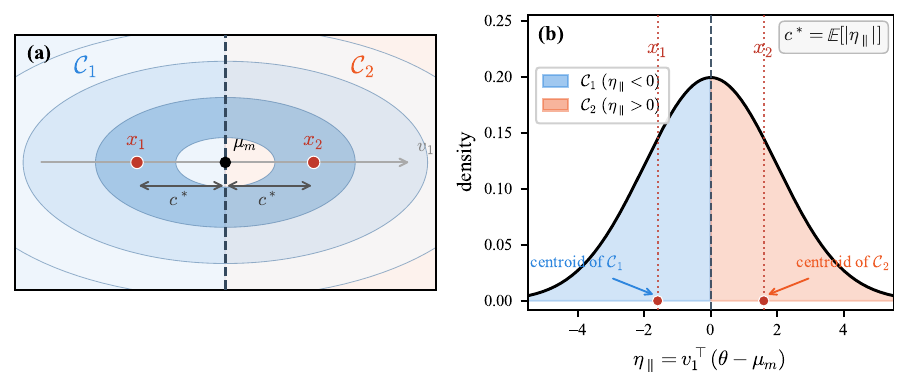}
\caption{Optimal two-product assortment under a bivariate Gaussian posterior ($d = 2$, $\lambda_1 = 4$, $\lambda_2 = 1$).} 
\label{fig:k2}
{\begin{flushleft}\footnotesize{\it Note.} (a)~Products $\bx_1$ and $\bx_2$ are placed symmetrically with respect to the posterior mean $\bmu_m$ along the leading eigenvector $\bv_1$, the direction of greatest uncertainty. The dashed decision boundary partitions the feature space into Voronoi regions $\cC_1$ and $\cC_2$, where each customer selects the nearer product. (b)~Marginal distribution of the customer's deviation $\eta_\parallel = \bv_1^\top(\btheta - \bmu_m)$ along the hedging direction. The optimal spread $c^* = \sqrt{2\lambda_1/\pi}$ places each product at the posterior centroid of its Voronoi region, i.e., $c^* = \E[|\eta_\parallel|]$.\end{flushleft}}
\end{figure}

\begin{example}[{\sc Two-Product Laptop Assortment}]\label{ex:k2}
Return to the travel laptop customer. Suppose that after solicitation, the posterior covariance has two principal directions: the portability-performance tradeoff (with remaining variance $\lambda_1 = 4$) and screen size (with $\lambda_2 = 1$). The agent is four times more uncertain about the customer's portability-performance preference than about screen size.

Because the portability-performance trade-off is where the agent knows the least, a second product helps the most. The agent recommends a lightweight ultrabook and a higher-performance model, placed on opposite sides of its best estimate along this axis. A customer who favors portability selects the ultrabook, whereas one who favors performance selects the other. Either way, the selected laptop is already on the correct side of the agent's best guess, so the remaining mismatch along this dimension is smaller than it would be with a single recommendation.

Both laptops, however, share the same screen size, as determined by the agent's best estimate. Offering two options does nothing to address screen-size uncertainty. If the customer prefers a larger or smaller screen, both laptops miss by the same amount. The only way to reduce screen-size mismatch is to ask more questions about screen preferences, rather than adding a second product.

How far apart should the two laptops be? Each one should be calibrated not to the average customer, but to its own target segment. The ultrabook is positioned at the average ideal point of portability-leaning customers, and the performance model at the average ideal point of performance-leaning customers, giving a spread of $c^* = \sqrt{2\lambda_1/\pi} \approx 1.60$. This diversification improves expected utility by $\lambda_1/\pi = 4/\pi \approx 1.27$, eliminating roughly $64\%$ of the expected loss along this dimension. Had the agent instead diversified along the screen-size axis, the gain would be only $\lambda_2/\pi \approx 0.32$, while leaving the larger portability-performance uncertainty entirely unaddressed. Figure~\ref{fig:k2} illustrates this geometry.\qed
\end{example}

Proposition~\ref{prop:k2-hedging} delivers two insights about how the agent should design its assortment pair. First, the agent should diversify in directions where it is most uncertain. The gain from a second product, $\lambda_1/\pi$, is proportional to the largest posterior variance, so the benefit depends on how concentrated the remaining uncertainty is. When uncertainty is spread roughly evenly across all dimensions ($\lambda_1 \approx \lambda_d$), no single direction stands out, and the gain from a second product is modest relative to the total loss. When uncertainty is sharply concentrated in one dimension ($\lambda_1 \gg \lambda_2$), a second product provides substantial improvement, because the direction it targets accounts for most of the remaining mismatch. With more than two products, the agent can diversify across multiple dimensions simultaneously, as formalized in Proposition~\ref{prop:rec-structure}.

Second, each product in the optimal pair is tailored to a customer segment rather than to the average customer. The two products implicitly divide customers into two groups: those who lean one way along the hedging direction and those who lean the other, and each product is placed at the average ideal point of its group. For larger assortments, the same centroid logic holds. We formalize the general properties in the next proposition.

\begin{proposition}[{\sc Optimal Multi-Product Assortment}]\label{prop:rec-structure}
Fix a history $\cH_m$ with posterior $\btheta \mid \cH_m \sim \mathcal{N}(\bmu_m, \bSigma_m)$, and let $\cS_k^*$ denote an optimal assortment of size $k$. Then:
\begin{enumerate}
    \item[(i)] \textbf{Centroid optimality:} $\cS_k^*$ solves the optimal $k$-point quantization problem
    $$
        \cS_k^* \in \arg\min_{\bx_1, \ldots, \bx_k \in \mathbb{R}^d} \E_{\btheta \sim \mathcal{N}(\bmu_m, \bSigma_m)}\left[\min_{j \in \{1,\ldots,k\}} \|\btheta - \bx_j\|_2^2\right],
    $$
    which partitions the feature space into $k$ Voronoi cells with each recommended product at the posterior centroid of its cell whenever that cell has positive posterior probability.
    \item[(ii)] \textbf{Symmetry of the objective:} If a set $\cS$ is optimal, then its reflection about $\bmu_m$, $2\bmu_m - \cS := \{2\bmu_m - \bx : \bx \in \cS\}$, is also optimal.
    \item[(iii)] \textbf{Distortion form:} Define the \emph{Gaussian $k$-point distortion}
    $$
        D_k(\bSigma) = \min_{\bx_1,\ldots,\bx_k\in\R^d} \E_{\boldsymbol{\xi}\sim\mathcal N(\boldsymbol{0},\bSigma)} \left[\min_{j\le k}\|\boldsymbol{\xi}-\bx_j\|_2^2\right].
    $$
    Then $V_k^*(\cH_m) = -\frac{1}{2}D_k(\bSigma_m)$. In particular, $V_k^*(\cH_m)$ depends on $\bSigma_m$ but not on $\bmu_m$.
\end{enumerate}
\end{proposition}

For general $k$, centroid optimality has the same partition-and-target interpretation as in the two-product case: the optimal assortment partitions the preference space into Voronoi regions and places each product at the centroid of its region.
The Gaussian posterior is centrally symmetric about $\bmu_m$, and, therefore, the assortment problem is invariant under reflection about $\bmu_m$. This implies that if a $k$-product set is optimal, then the mirrored set $2\boldsymbol{\mu}_m - \mathcal{S}$ is also optimal. Hence, there is no systematic directional bias in the assortment objective.

The distortion form provides a unified expression for the terminal value across all $k$. The distortion $D_k(\bSigma)$ measures how effectively $k$ optimally placed products can cover posterior uncertainty: for $k=1$, $D_1(\bSigma) = \tr(\bSigma)$ recovers the trace formula of Proposition~\ref{prop:single-rec}, while larger $k$ exploit richer eigenvalue structure. The mean-invariance of $V_k^*$ generalizes the same property from the single-product case and carries a far-reaching consequence for the solicitation phase: the agent's solicitation policy should be chosen to minimize $D_k(\bSigma_m)$, regardless of how the posterior mean evolves during solicitation. Let $V_\infty^* := \lim_{k\to\infty} V_k^*$ denote the full-information benchmark.

\begin{proposition}[{\sc Value of Assortment Breadth}]\label{prop:assortment}
For any realization of $\cH_m$, the optimal expected utility satisfies:
\begin{enumerate}
    \item[(i)] $V_k^*$ is increasing in $k$.
    \item[(ii)] $V_\infty^* = 0$.
    \item[(iii)] $V_\infty^* - V_k^* = O(k^{-2/d})$ as $k\to\infty$.
\end{enumerate}
\end{proposition}

Expanding the recommendation assortment improves expected match quality because customers select the product closest to their ideal point. As shown in Proposition~\ref{prop:k2-hedging}, the first expansion from one to two products captures a gain of $\lambda_1/\pi$, where $\lambda_1$ is the largest eigenvalue of $\bSigma_m$, by hedging along the direction of greatest posterior uncertainty. As $k$ grows, the remaining gap $V_\infty^* - V_k^*$ shrinks at a rate of $O(k^{-2/d})$. In the limit, the expected loss vanishes, achieving the full-information benchmark $V_\infty^* = 0$. The gap between the full-information benchmark and the single-recommendation expected utility,
\begin{equation}\label{eq:hedge-value}
    V_\infty^* - V_1^* = \frac{1}{2}\tr(\bSigma_m),
\end{equation}
quantifies the total potential gain from assortment breadth. When posterior uncertainty is low, even a single recommendation performs well, leaving little room for improvement by adding additional products. When posterior uncertainty remains large, greater assortment breadth yields larger gains.

Although the formal model treats $k$ as given, this result suggests that the agent need not commit to a fixed assortment size. Instead, it can adapt $k$ to the realized posterior uncertainty. When $\tr(\bSigma_m)$ is small, a single well-placed recommendation nearly attains the full-information benchmark, and presenting fewer options reduces the customer's cognitive burden, consistent with concerns about choice overload \citep{IyengarLepper2000}. When $\tr(\bSigma_m)$ remains large, offering additional products allows the assortment to hedge against the residual uncertainty. This adaptive sizing is a practical advantage of the agentic purchasing paradigm: the same posterior covariance that guides \emph{what} to recommend also determines \emph{how many} products to present. The dependence on $\bSigma_m$ creates a natural link to the solicitation policy, which we explore in Section~\ref{sec:interplay}.

\subsection{Optimal Solicitation Policy}\label{sec:solicitation}

We now turn to the solicitation problem: how should the agent choose its query directions $(\by_1, \ldots, \by_m)$? We begin with the single-product assortment benchmark ($k=1$), which admits an explicit characterization and reveals the core economic logic. We extend the analysis to general $k$ after developing the benchmark.

Specifically, when $k=1$, Proposition~\ref{prop:single-rec} gives terminal utility $V_1^* = -\frac{1}{2}\tr(\bSigma_m)$, which depends on the posterior covariance alone. The solicitation problem, therefore, reduces to choosing query directions that minimize $\tr(\bSigma_m)$. To formalize this, for $t = 0, 1, \ldots, m-1$, define the value-to-go
$J_t(\cH_t) = \max_{\pi_{t+1}, \ldots, \pi_m, \rho} \E[ V(\rho(\cH_m); \btheta) \,|\, \cH_t ]$ satisfying the Bellman recursion
$$
    J_t(\cH_t) = \max_{\by : \|\by\|_2 = 1} \E\left[ J_{t+1}(\cH_t, \by, z_{t+1}) \,\Big|\, \cH_t, \by \right],
$$
with boundary condition $J_m(\cH_m) = -\frac{1}{2}\tr(\bSigma_m)$ from Proposition~\ref{prop:single-rec}.
As the covariance update~\eqref{eq:cov-update} depends on $\by_t$ but not on the realized response $z_t$, the terminal objective depends on the history only through the realized query sequence. Therefore, the dynamic program reduces to
\begin{equation}\label{eq:k1-solicitation}
    \min_{\by_1, \ldots, \by_m : \|\by_t\|_2 = 1} \tr(\bSigma_m).
\end{equation}
This trace-minimization objective coincides with the classical A-optimality criterion in experimental design \citep{Pukelsheim2006}, but here it emerges endogenously from the economic model rather than being imposed as a design surrogate.

\begin{lemma}[{\sc Solicitation Gains for $k=1$}]\label{lem:solicitation}
When the agent offers one product, the solicitation seeks to reduce $\tr(\bSigma_m)$. Moreover:
\begin{enumerate}
    \item[(i)] \textbf{Per-round gain and optimal direction.} At round $t$ with posterior covariance $\bSigma_{t-1}$, querying any unit vector $\by$ yields utility gain
    $$
        \Delta_t(\by) = \frac{1}{2}\,\frac{\by^\top \bSigma_{t-1}^2 \by}{\sigma^2 + \by^\top \bSigma_{t-1} \by},
    $$
    which is maximized by any leading eigenvector of $\bSigma_{t-1}$.
    \item[(ii)] \textbf{Telescoping identity.} For any query sequence $(\by_1,\ldots,\by_m)$,
    \begin{equation}\label{eq:vos-formula}
        \textup{VOS}(m) := \sum_{t=1}^{m} \Delta_t(\by_t) = \frac{1}{2}\left( \tr(\bSigma_0) - \tr(\bSigma_m) \right).
    \end{equation}
\end{enumerate}
\end{lemma}

Lemma \ref{lem:solicitation} part (i) shows that a single question is most valuable when it targets the direction of greatest posterior uncertainty. Part (ii) gives path-independent accounting: for $k=1$, the cumulative value of solicitation depends on the query path only through the terminal covariance $\bSigma_m$ via \eqref{eq:vos-formula}. This identity will anchor the solicitation-assortment trade-off in Section~\ref{sec:interplay}.

The per-round greedy rule of always querying the leading eigenvector does not by itself determine the optimal $m$-round policy, because the agent must allocate a finite solicitation budget across dimensions. The following proposition shows that the optimal allocation follows a water-filling equalization in the prior eigenbasis, subject to the rank cap imposed by $m$ rank-one queries.

\begin{proposition}[{\sc Optimal Solicitation Policy for $k=1$: Rank-Capped Water-Filling}]\label{prop:k1-rankcap}
Let $\bSigma_0=\sum_{i=1}^d \lambda_i(0)\,\bv_i\bv_i^\top$ with $\lambda_1(0)\ge\cdots\ge\lambda_d(0)>0$. The $k=1$ solicitation problem~\eqref{eq:k1-solicitation} admits an optimal policy whose aggregate information matrix $\boldsymbol{M}_m=\sum_{t=1}^m \by_t\by_t^\top$ is diagonal in the prior eigenbasis $\{\bv_i\}$. Equivalently, the optimal posterior covariance is diagonal in this eigenbasis with eigenvalues
$$
\lambda_i^\star(m)=
\begin{cases}
\lambda^\star, & i=1,\ldots,r^\star,\\
\lambda_i(0), & i=r^\star+1,\ldots,d,
\end{cases}
$$
where $r^\star\le \min\{m,d\}$ is the number of directions that receive at least one question and $\lambda^\star$ is the common posterior variance across those directions, satisfying
$$
\frac{r^\star}{\lambda^\star}=\sum_{i=1}^{r^\star}\frac{1}{\lambda_i(0)}+\frac{m}{\sigma^2},
\qquad
\lambda^\star< \lambda_{r^\star}(0),
$$
and, if $r^\star<\min\{m,d\}$,
$$
\lambda^\star\ge \lambda_{r^\star+1}(0).
$$
Any directions tied at the boundary, with $\lambda_i(0)=\lambda^\star$, can remain unqueried without affecting the optimal posterior covariance.
\end{proposition}

Proposition~\ref{prop:k1-rankcap} yields a rank-capped water-filling rule. The agent allocates its $m$ questions to the directions whose prior variances exceed the common level $\lambda^\star$, equalizing those posterior variances at $\lambda^\star$, and leaves the remaining directions unqueried at their prior variances. The characterization is naturally expressed in terms of the aggregate information matrix $\boldsymbol{M}_m$, which records the distribution of the $m$ questions across preference directions. An explicit $m$-round query sequence that implements the optimal $\boldsymbol{M}_m$ is given in the Online Appendix (see Lemma~EC.2 and the proof of Proposition~\ref{prop:k1-rankcap}).

To identify which directions receive questions, consider for each $r\le \min\{m,d\}$ the equalization level
$$
\lambda(r)=\frac{r}{\sum_{i=1}^{r}\lambda_i(0)^{-1}+m/\sigma^2}.
$$
Then $r^\star$ is the unique index satisfying $\lambda_{r^\star}(0)>\lambda(r^\star)$ and, if $r^\star<\min\{m,d\}$, $\lambda(r^\star)\ge \lambda_{r^\star+1}(0)$.
The left inequality ensures that the agent queries direction $r^\star$, while the right inequality indicates that querying the next direction would not improve the objective.

Each question targets one direction in the preference space, so $m$ questions can address at most $m$ independent directions, giving the cap $r^\star\le m$. The water-filling condition equalizes the marginal value of an additional question across all queried dimensions at the common variance $\lambda^\star$. As $m$ increases, $\lambda^\star$ declines and additional directions begin to receive questions, eventually yielding $r^\star=d$ and uniform posterior uncertainty across all dimensions once the budget is large enough.

When the agent recommends $k \ge 2$ products, assortment breadth can hedge against residual uncertainty, so solicitation must consider not only how much to learn but also how to shape what remains. We illustrate this tradeoff with $k=2$, where the distortion $D_2$ admits a closed form.

\begin{example}[{\sc Two-Product Solicitation Policy}]\label{ex:k2-solicitation}
Consider $k = 2$. Proposition~\ref{prop:k2-hedging} implies that for any covariance matrix $\bSigma$ with eigenvalues $\lambda_1 \ge \cdots \ge \lambda_d$,
$$
D_2(\bSigma) = \tr(\bSigma) - \frac{2}{\pi}\lambda_1.
$$
The solicitation objective therefore becomes $\min\, \tr(\bSigma_m) - \frac{2}{\pi}\lambda_1(\bSigma_m)$. Compared to the $k=1$ case of pure trace minimization, the additional term $-\frac{2}{\pi}\lambda_1(\bSigma_m)$ rewards preserving a single direction of high residual uncertainty, an asymmetry that the assortment can exploit through hedging. The solicitation policy, therefore, faces a tradeoff absent in the single-product case: whether to reduce total uncertainty or to concentrate the remaining uncertainty to benefit the assortment stage. Return to the travel laptop customer, and suppose the agent is equally uncertain about the customer's preferred screen size and portability-performance tradeoff. When recommending a single laptop ($k=1$), the agent spreads its questions evenly across both dimensions: all uncertainty is costly, and the best strategy is to learn as much as possible. When recommending two laptops ($k=2$), the calculation changes. The agent can pair an ultrabook with a performance model to hedge along the portability-performance axis. Residual uncertainty along this axis is no longer purely a cost: wherever the customer falls on this trade-off, one of the two laptops is a close match. When the solicitation budget is large enough for this multi-product trade-off to emerge, the optimal solicitation asks fewer questions about the portability-performance trade-off and more about screen size, selectively focusing precision on screen size and leaving the uncertainty that the two-product assortment can exploit. By contrast, when the agent has fewer questions than the preference dimensionality, it cannot actively learn all dimensions, and this tradeoff need not arise. Under an isotropic prior, Lemma~\ref{lem:k2-solicitation}(i) shows that the $k=2$ optimum coincides with the $k=1$ rank-capped water-filling benchmark in this regime. \qed
\end{example}

Under an isotropic prior, the tradeoff in Example~\ref{ex:k2-solicitation} admits an explicit characterization.

\begin{lemma}[{\sc Two-Product Solicitation under Isotropic Prior}]\label{lem:k2-solicitation}
Suppose $d\ge 2$, $\bSigma_0=\sigma_0^2\bI$, and $k=2$, so $D_2(\bSigma)=\tr(\bSigma)-\frac{2}{\pi}\lambda_1(\bSigma)$. Let $\gamma=\sqrt{1-2/\pi} \approx 0.603$.
\begin{enumerate}
\item[(i)] \textbf{Selective focus.} Let $\bSigma_m^{\star,2}$ be a $k=2$ optimal posterior and write its eigenvalues as $\lambda_1^*\ge\cdots\ge\lambda_d^*$. Then $\bSigma_m^{\star,2}$ has at most two distinct eigenvalues. When $m < d$, $\lambda_1^*=\cdots=\lambda_{d-m}^*=\sigma_0^2$ and $\lambda_{d-m+1}^*=\cdots=\lambda_d^*$. When $m \ge d$, $\lambda_1^* \ge \lambda_2^* = \cdots = \lambda_d^*$. Let $\hat{m} = \frac{1-\gamma}{\gamma}(d-1)\frac{\sigma^2}{\sigma_0^2}$. When $m < d$, the $k=1$ rank-capped water-filling policy (Proposition~\ref{prop:k1-rankcap}) is optimal for $k=2$. For $d \le m \le \hat{m}$, the optimal $k=2$ policy keeps $\lambda_1^* = \sigma_0^2$ and concentrates all precision on the remaining $d-1$ directions, yielding
$$
\frac{\lambda_1^*}{\lambda_2^*} = 1 + \frac{m\,\sigma_0^2}{(d-1)\,\sigma^2},
$$
which grows continuously toward $1/\gamma$ as $m$ increases. For $m \ge d$ and $m > \hat{m}$, the ratio locks at the universal constant
$$
\frac{\lambda_1^*}{\lambda_2^*} = \frac{1}{\gamma} \approx 1.659.
$$
\item[(ii)] \textbf{Near-optimality of the single-product benchmark.} Let $\bSigma_m^{\star,1}$ denote the rank-capped water-filling posterior from Proposition~\ref{prop:k1-rankcap}, and let $\bSigma_m^{\star,2}$ be a $k=2$-optimal posterior. Define
$$
R_2(d,m) \;:=\; \frac{D_2(\bSigma_m^{\star,1})}{D_2(\bSigma_m^{\star,2})}.
$$
If $m<d$, then $\bSigma_m^{\star,1}$ is also $k=2$ optimal, so $R_2(d,m)=1$. Moreover, for all $m$,
$$
R_2(d,m) \;\le\; \frac{d\,(d - 2/\pi)}{(\gamma + d - 1)^2},
$$
and equality holds whenever $m \ge d$ and $m > \hat{m}$. The upper bound gives $R_2(2,m) \le 1.061$, $R_2(5,m) \le 1.030$, and $R_2(d,m) - 1 = O(1/d)$ as $d \to \infty$.
\end{enumerate}
\end{lemma}

When $m < d$, the rank constraint limits the policy to at most $m$ queried directions, so the $k=1$ rank-capped water-filling policy is also optimal for $k=2$. Once $m\ge d$, the two policies diverge. The $k=1$ policy equalizes all posterior eigenvalues, but in the transitional regime (i.e., $d\le m\le \hat{m}$), the $k=2$ optimal policy asks no questions about the leading direction, leaving its posterior variance at the prior level $\sigma_0^2$, and devotes all questions to the remaining $d-1$ directions. The reason is that the two-product assortment will hedge along the leading direction, absorbing a fraction $2/\pi$ of the variance there: each unit of residual variance in this direction costs only $1 - 2/\pi$ units of distortion rather than a full unit, so each additional question about that direction reduces distortion by less than the same question asked about other directions. As $m$ increases within this regime, the ratio of the largest to smallest posterior eigenvalue, $\lambda_1^*/\lambda_2^*$, grows continuously until it reaches $1/\gamma \approx 1.659$ at $m = \hat{m}$. Once the number of solicitation rounds reaches $\hat{m}$, the per-question distortion reduction in the hedging direction, after accounting for the variance that hedging absorbs, finally equals the per-question reduction in the remaining directions, so the optimal policy begins to ask questions about all directions. The eigenvalue ratio $\lambda_1^*/\lambda_2^*$ then stabilizes at the universal constant $1/\gamma$, which depends solely on the geometric constant $2/\pi$ and is invariant to all problem parameters. This selective focus reflects the core tradeoff in multi-product solicitation: whether to reduce total variance or to concentrate residual variance along a direction that the assortment can exploit through hedging.

Despite this qualitative departure from the single-product case, the distortion penalty from using the simpler $k=1$ benchmark, rank-capped water-filling, remains uniformly small. The near-optimality bound in Lemma~\ref{lem:k2-solicitation} shows that $R_2(d,m)=1$ when $m<d$ and that for every $m\ge d$, $R_2(d,m)$ is bounded above by a dimensionality-dependent-only constant, which is at most $6.1\%$ when $d=2$ and satisfies $1+O(1/d)$ as $d\to\infty$. The benchmark posterior $\bSigma_m^{\star,1}$ is induced by a feasible $m$-round query sequence, so it is feasible for the $k=2$ solicitation problem by construction. This suggests that using the $k=1$ capped water-filling solicitation policy incurs only a small loss even when the assortment stage offers two products. As the next result shows, the $k=1$ policy admits a general approximation guarantee for all $k$.

For general $k$, Proposition~\ref{prop:rec-structure}(iii) gives $V_k^*(\cH_m) = -\frac{1}{2}D_k(\bSigma_m)$, so the solicitation problem becomes
$$
\min_{\by_1,\ldots,\by_m:\; \|\by_t\|_2=1}
D_k(\bSigma_m),
\qquad
\text{s.t.}\quad
\bSigma_m^{-1}=\bSigma_0^{-1}+\frac{1}{\sigma^2}\sum_{t=1}^m \by_t\by_t^\top,
$$
where the constraint follows from iterating the Kalman update (Lemma~\ref{lem:posterior}). The tractability of the $k=1$ case relied on the trace objective $D_1(\bSigma) = \tr(\bSigma)$ being separable in eigenvalues. For general $k$, $D_k$ is not separable and does not admit a closed-form expression, so the resulting optimization problem is generally nonconvex even under Gaussian priors and continuous-allocation relaxations.
Nevertheless, the following proposition provides an approximation guarantee for using the $k=1$ capped water-filling policy for any $k$.
As a benchmark for quantization quality, define the \emph{univariate quantization efficiency}
$$
\eta_k \;=\; 1 \;-\; \frac{\min_{c_1,\ldots,c_k}\, \E\bigl[\min_i (X - c_i)^2\bigr]}{\E[X^2]}, \qquad X \sim \mathcal{N}(0,1),
$$
the fraction of variance eliminated by optimally approximating a standard normal variable with $k$ points. The constant $2/\pi$ in Lemma~\ref{lem:k2-solicitation} is precisely $\eta_2$. For larger $k$, $\eta_k$ can be computed via the Lloyd-Max algorithm. Appendix~A describes the procedure, and Table~A.1 reports values for representative $k$.

\begin{proposition}[{\sc Capped Water-Filling Approximation Guarantee}]\label{pro:wf-approx}
Let $\bSigma_m^{\star,1}$ denote the rank-capped water-filling posterior from Proposition~\ref{prop:k1-rankcap}, and for $k \ge 2$ let $\bSigma_m^{\star,k}$ denote an optimal posterior for $\min D_k(\bSigma_m)$. Define the \emph{isoperimetric ratio}
$$
\alpha(\bSigma) \;:=\; \frac{\tr(\bSigma)}{d\,(\det\bSigma)^{1/d}} \;\ge\; 1,
$$
with equality if and only if $\bSigma$ is isotropic. For any Gaussian prior covariance $\bSigma_0 \succ 0$ and all $k \ge 2$,
$$
\frac{D_k(\bSigma_m^{\star,1})}{D_k(\bSigma_m^{\star,k})} \;\le\; \alpha\!\left(\bSigma_m^{\star,1}\right)\, k^{2/d},
$$
where $\alpha(\bSigma_m^{\star,1})$ is convex and non-increasing in $m$ for $m \ge d$ and equals $1$ whenever $m \ge \max\{d,\bar{m}\}$, where $\bar{m} \,:=\, \sigma^2\!\bigl(\frac{d}{\lambda_{\min}(\bSigma_0)} - \tr(\bSigma_0^{-1})\bigr)^{+}$.
\end{proposition}

Proposition~\ref{pro:wf-approx} establishes an approximation bound for using the capped water-filling solicitation policy designed for a single-product assortment with any $k$. The approximation ratio can be decomposed into two interpretable factors.
The isoperimetric ratio $\alpha(\bSigma_m^{\star,1}) \ge 1$ measures how unevenly residual uncertainty is spread across preference dimensions after solicitation, with equality when uncertainty is uniform across all directions. As the solicitation progresses, the $k=1$ capped water-filling policy (Proposition~\ref{prop:k1-rankcap}) equalizes the residual uncertainty, thereby driving the isoperimetric ratio toward one. Once $m \ge d$, so that every preference dimension has been queried at least once, the isoperimetric ratio begins to decline toward one with each additional question, and the convexity of $\alpha$ in $m$ ensures that the steepest improvement occurs in the earliest rounds after this point. By the time $m$ reaches $\max\{d,\bar{m}\}$, the isoperimetric ratio equals one exactly and the approximation ratio reduces to the geometric penalty $k^{2/d}$ alone. When the preference dimensionality is not too large, this threshold is reached quickly, so a few rounds of solicitation suffice to make the simple $k=1$ water-filling policy a good approximation for any fixed $k$.

The geometric penalty $k^{2/d}$ reflects the cost of covering a $d$-dimensional uncertainty region with only $k$ discrete products. When only a few products are recommended, this factor remains bounded, and the substitutability result below (Proposition~\ref{prop:substitutes}) provides the economic rationale for keeping $k$ small. Given its reduction in the isoperimetric ratio due to solicitation, the capped water-filling policy for a single product can be a useful benchmark across assortment sizes that remain modest relative to the dimensionality.

\subsection{Solicitation and Assortment Interplay}\label{sec:interplay}

The preceding analysis optimizes one instrument at a time: Section~\ref{sec:recommendation} characterizes the optimal assortment for a given posterior, and Section~\ref{sec:solicitation} designs the solicitation policy anticipating a fixed assortment size $k$. We now endogenize both instruments simultaneously and examine their interaction. The agent can improve expected utility in two ways: by asking more questions (increasing $m$) during the solicitation phase, or by offering more choices (increasing $k$) during the assortment phase. Both address the same source of loss: residual uncertainty about the customer's ideal product, and both draw on a common pool of prior uncertainty. The combination of \eqref{eq:hedge-value} with \eqref{eq:vos-formula} yields the \emph{uncertainty decomposition identity}:
$$
    \textup{VOS}(m) + (V_\infty^* - V_1^*) = \frac{1}{2}\tr(\bSigma_0).
$$
The total prior uncertainty $\frac{1}{2}\tr(\bSigma_0)$ is a fixed pie. The first term quantifies the uncertainty eliminated by solicitation depth, and the second quantifies the maximum remaining uncertainty that assortment breadth can hedge. Each additional question shifts value from the second term to the first, making the two instruments substitutes.
Moreover, each question increases total precision by exactly $1/\sigma^2$, so the posterior satisfies the precision budget identity $\tr(\bSigma_m^{-1})=\tr(\bSigma_0^{-1})+m/\sigma^2$. Since the hedging gap equals $\frac{1}{2}\tr(\bSigma_m)$, this precision budget directly governs how fast solicitation depth shrinks the residual uncertainty available for assortment breadth to hedge. The combination of this solicitation rate with the breadth rate from Proposition~\ref{prop:assortment}(iii) delivers a tight comparison between solicitation depth and assortment breadth.

\begin{proposition}[{\sc Solicitation-Assortment Substitutability}]\label{prop:substitutes}
Consider the Gaussian model and let $\bSigma_m$ denote the posterior covariance after $m$ questions under any solicitation policy.
\begin{enumerate}
    \item[(i)] \textbf{Solicitation rate.} The hedging gap satisfies
    $$
        V_\infty^* - V_1^* \;\ge\; \frac{d^2}{2(\tr(\bSigma_0^{-1}) + m/\sigma^2)}.
    $$
    The rank-capped water-filling policy (described in Proposition~\ref{prop:k1-rankcap}) attains this bound once $m \ge \max\{d,\bar{m}\}$, where $\bar{m}$ is the activation threshold from Proposition~\ref{pro:wf-approx}.
    \item[(ii)] \textbf{Substitutability.} Fix $\varepsilon>0$. Achieving $V_\infty^* - V_1^* \le \varepsilon$ through solicitation requires
    $$
        m \;\ge\; \sigma^2\!\left(\frac{d^2}{2\varepsilon}-\tr(\bSigma_0^{-1})\right)^{+}
    $$
    under any policy. Once $m \ge \max\{d,\bar{m}\}$, the rank-capped water-filling policy attains the bound in part~(i), so for fixed $\bSigma_0$ this implies $m=\Theta(d^2\sigma^2/\varepsilon)$ as $\varepsilon\downarrow 0$.
    By contrast, without solicitation ($m=0$), achieving $V_\infty^* - V_k^* \le \varepsilon$ through assortment breadth alone requires
    $$
        k \;\ge\; \left(\frac{d(\det \bSigma_0)^{1/d}}{2\varepsilon}\right)^{d/2},
    $$
    and hence $k=\Omega(\varepsilon^{-d/2})$ for fixed $\bSigma_0$ as $\varepsilon\downarrow 0$.
\end{enumerate}
\end{proposition}

Proposition~\ref{prop:substitutes} reveals a fundamental rate asymmetry between the agent's two instruments. Each question yields a fixed precision gain of $1/\sigma^2$, so no query strategy, however adaptive, can close the hedging gap faster than at a rate $1/m$. For a fixed prior covariance $\bSigma_0$, to drive $V_\infty^* - V_1^*$ below a small tolerance $\varepsilon$ therefore requires on the order of $d^2\sigma^2/\varepsilon$ questions, and rank-capped water-filling attains this rate once every preference dimension has been queried. By contrast, recommendation breadth faces the geometric penalty of covering a $d$-dimensional uncertainty region with $k$ points, which forces the distortion gap to decay at rate $k^{-2/d}$. Without solicitation, achieving $V_\infty^* - V_k^* \le \varepsilon$ requires $k=\Omega(\varepsilon^{-d/2})$ products for fixed $\bSigma_0$. This polynomial-versus-exponential gap implies that a single well-targeted question can be worth more than adding another product to the assortment.

Because of this rate asymmetry, even moderate solicitation renders large assortments unnecessary. Two forces drive this. First, after a few targeted questions, residual uncertainty concentrates along a handful of preference dimensions, so two or three products hedging along those directions capture most of the achievable gain (Proposition~\ref{prop:k2-hedging}). Second, after extensive solicitation, most uncertainty has been resolved (Proposition~\ref{prop:k1-rankcap}), which leaves little for any number of products to exploit. The choice overload literature provides additional behavioral support: excessive options impair decision quality and customer satisfaction \citep{IyengarLepper2000}, and the substitutability result provides a formal economic rationale for limiting the size of the recommendation assortment.

Effective solicitation reinforces this substitutability through a second channel. Beyond shrinking total uncertainty, the capped water-filling policy (see Proposition~\ref{prop:k1-rankcap}) equalizes the remaining uncertainty across all queried dimensions, eliminating the concentration of uncertainty along any single direction that a large product set would otherwise exploit through hedging. Assortment breadth is most effective when uncertainty is spectrally concentrated. For $k=2$, $D_2(\bSigma)=\tr(\bSigma)-\frac{2}{\pi}\lambda_1(\bSigma)$ makes this explicit. For fixed $\tr(\bSigma)$, an increase in $\lambda_1(\bSigma)$ lowers $D_2(\bSigma)$, so equalization of eigenvalues weakens the need for leveraging the recommendation assortment breadth. Capped water-filling, therefore, compounds two effects. It reduces not only the total posterior variance but also the spectral concentration, both of which shrink the incremental value of a larger $k$. This contrasts with the intuition that a traditional retailer without individual solicitation must rely on broad assortments to accommodate heterogeneity in preferences across the entire customer population. Here, personalized solicitation replaces population-level coverage and resolves individual uncertainty through solicitation depth rather than assortment breadth.

\begin{figure}[htp]
\centering
\includegraphics[width=\textwidth]{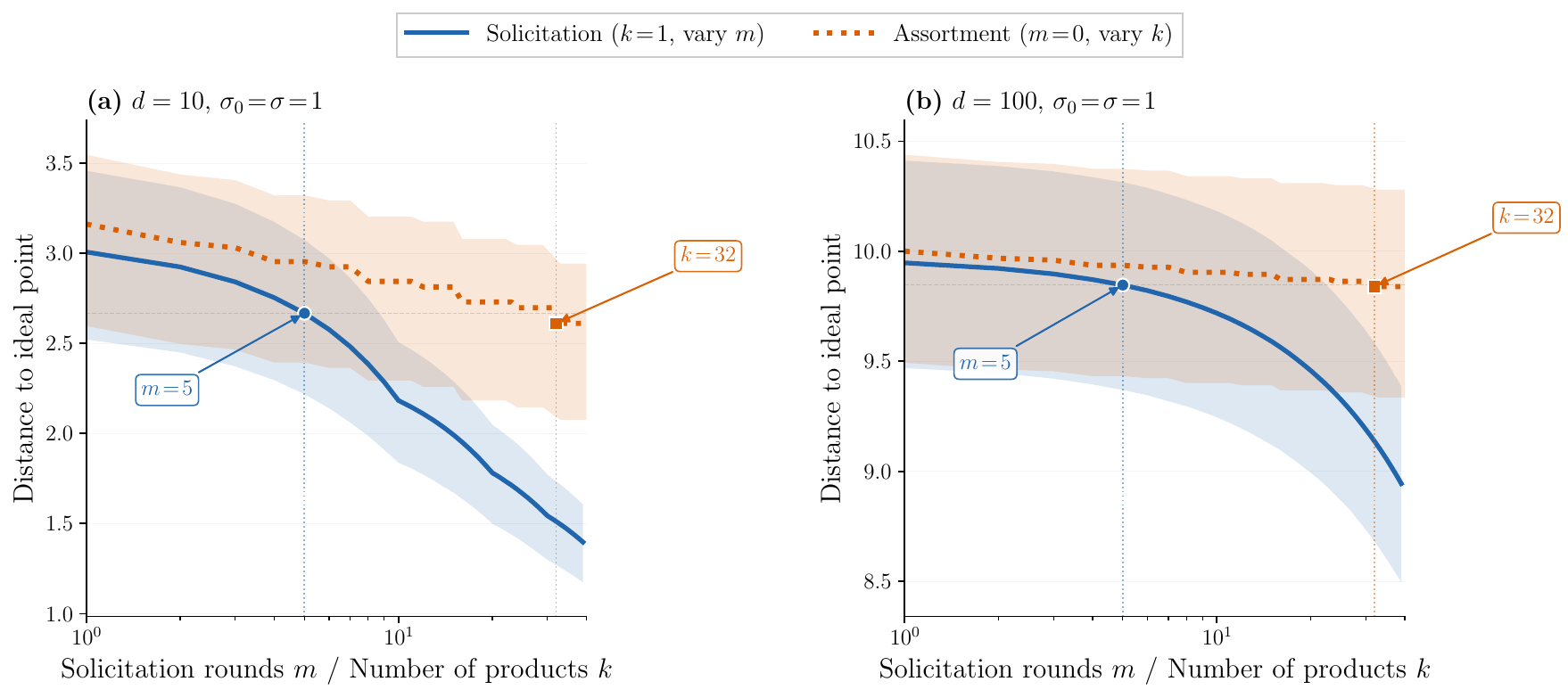}
\caption{Comparison of solicitation depth versus assortment breadth.}
\label{fig:simulation}
{\begin{flushleft}\footnotesize{\it Note.} Blue solid line: expected distance when varying the number of questions $m$ with a single recommendation ($k = 1$), using rank-capped water-filling (Proposition~\ref{prop:k1-rankcap}). Orange dotted line: expected distance when varying the number of products $k$ with no solicitation ($m = 0$), using a product quantizer constructed from Lloyd-Max centroids. Lines display sample means, and shaded bands span the 25th to 75th percentiles. Both axes use a logarithmic scale.\end{flushleft}}
\end{figure}

Figure~\ref{fig:simulation} illustrates that asking questions is far more effective than expanding the assortment. We simulate $50{,}000$ customer ideal points drawn from an isotropic Gaussian prior $\btheta \sim \mathcal{N}(\boldsymbol{0}, \sigma_0^2 \bI)$ with $\sigma_0 = \sigma = 1$, and compare two strategies for reducing the expected distance between the recommended product and the customer's ideal point. The first strategy asks $m$ questions while recommending a single product, using the rank-capped water-filling policy (Proposition~\ref{prop:k1-rankcap}). The second strategy offers $k$ products without any solicitation, placing them via Lloyd-Max centroids. Each strategy starts from the same uninformed baseline ($m = 0$, $k = 1$) and varies only one instrument, isolating its individual contribution.

In both a moderate-dimensional setting ($d = 10$, panel~a) and a high-dimensional setting ($d = 100$, panel~b), five well-chosen questions with a single recommendation achieve roughly the same expected distance as an assortment of 32 products offered without solicitation. The underlying mechanism is the rate asymmetry established in Proposition~\ref{prop:substitutes}: to reach a given distance target, the number of questions the agent needs grows only polynomially in the dimensionality $d$, whereas the number of products grows exponentially. Because of this gap, more questions sharply reduce the distance to the customer's ideal point, while more products barely help, especially in high dimensions. Panel~(b) makes this concrete: with $d = 100$, the assortment curve is nearly flat as $k$ grows from 1 to 32, because each additional product covers only a negligible fraction of the 100-dimensional preference space. By contrast, additional questions continue to reduce distance effectively, because the total number of questions needed to reach a given accuracy is polynomial in $d$.

Together, these results yield a clear design prescription for AI-assisted purchasing: invest in solicitation to learn preferences, then present a small, targeted assortment. Moderate solicitation captures most of the achievable gain, making very large assortments economically unnecessary, and the advantage of questions over products widens as the number of relevant preference dimensions increases.

\section{Beyond Gaussian Priors}\label{sec:non-gaussian}

The Gaussian prior yields closed-form solutions, thereby making the economic forces transparent. We now extend the analysis to general priors.
Let $\btheta$ follow a general prior distribution $P_0$ on $\R^d$ with density $p_0$. Throughout this section, we maintain the following assumption.

\begin{assumption}\label{as:second-moment}
The prior $P_0$ satisfies $\E[\|\btheta\|^2] < \infty$, with mean $\bmu_0 = \E[\btheta]$ and covariance $\bSigma_0 = \text{Cov}(\btheta) \succ 0$.
\end{assumption}

As in the Gaussian case, at each round $t$, the agent selects a query direction $\by_t$ and observes $z_t = \btheta^\top \by_t + \epsilon_t$ with i.i.d. noise $\epsilon_t \sim \mathcal{N}(0, \sigma^2)$. We continue to write $\cH_t$ for the information available after round $t$. Since the prior $P_0$ and its moments $(\bmu_0,\bSigma_0)$ are fixed, we suppress them in the notation and let $\cH_t$ record the solicitation transcript $(\by_1,z_1,\ldots,\by_t,z_t)$. The posterior after history $\cH_t$ is
$$
p_t(\btheta \mid \cH_t) \propto p_0(\btheta) \prod_{s=1}^{t} \phi\!\left(\frac{z_s - \btheta^\top \by_s}{\sigma}\right),
$$
where $\phi$ denotes the standard normal density. The following result characterizes optimal solicitation and assortment under a general prior. Let $P_t = P(\btheta \mid \cH_t)$ denote the posterior after $t$ rounds, $\bthetahat_m = \E[\btheta \mid \cH_m]$ the posterior mean, and $\bSigma_m(\cH_m) = \textup{Cov}(\btheta \mid \cH_m)$ the posterior covariance after $m$ solicitation rounds.

\begin{proposition}[{\sc General Prior Analysis}]\label{prop:general-structure}
Under Assumption~\ref{as:second-moment}, we have: 
\begin{enumerate}
\item[(i)] \textbf{Value decomposition.} The optimal single-product assortment is the posterior mean $\bthetahat_m$, with terminal value $V_1^*(\cH_m) = -\frac{1}{2}\tr(\bSigma_m(\cH_m))$. The expected terminal value satisfies
$$
\E[V_1^*] = -\frac{1}{2}\tr(\bSigma_0) + \textup{VOS}(m), \quad \textup{VOS}(m) = \frac{1}{2}\tr(\textup{Cov}(\bthetahat_m)),
$$
where $\textup{Cov}(\bthetahat_m) = \textup{Cov}(\E[\btheta \mid \cH_m])$ is the covariance of the posterior mean.

\item[(ii)] \textbf{Monotonicity and uncertainty decomposition.} $\textup{VOS}(m)$ is increasing in $m$. The value of assortment breadth $\textup{VOA}(k; m) := \E[V_k^* - V_1^*]$ is increasing in $k$ and satisfies
$$
0 \le \textup{VOA}(k; m) \le \frac{1}{2}\E[\tr(\bSigma_m(\cH_m))] = \frac{1}{2}\tr(\bSigma_0) - \textup{VOS}(m).
$$
In particular, $\textup{VOS}(m) + \textup{VOA}(k; m) \le \frac{1}{2}\tr(\bSigma_0)$ and the upper bound on $\textup{VOA}(k; m)$ is decreasing in $m$.

\item[(iii)] \textbf{Conservative benchmark.} Fix any non-adaptive query directions $(\by_1,\ldots,\by_m)$ with $\|\by_t\|_2=1$. Let $\bSigma_m^G$ denote the posterior covariance under the Gaussian prior $\mathcal{N}(\bmu_0, \bSigma_0)$ after these $m$ queries. Then
$$
\E[\tr(\bSigma_m(\cH_m))] \leq \tr(\bSigma_m^G),
$$
and consequently $\textup{VOS}(m) \geq \textup{VOS}^G(m)$, where $\textup{VOS}^G(m) := \frac{1}{2}(\tr(\bSigma_0) - \tr(\bSigma_m^G))$ is the value of solicitation under the Gaussian benchmark.
\end{enumerate}
\end{proposition}

The value decomposition in Proposition~\ref{prop:general-structure}(i) separates the expected payoff into two terms: a baseline loss $-\frac{1}{2}\tr(\bSigma_0)$ from recommending under prior uncertainty alone, and the gain $\textup{VOS}(m)$ from solicitation. The gain equals half the trace of $\textup{Cov}(\bthetahat_m)$, which captures how much the posterior mean varies across customers. When this variation is large, solicitation leads to substantially personalized recommendations. When it is small, most customers receive similar recommendations regardless of their answers. The monotonicity and substitutability result in Proposition~\ref{prop:general-structure}(ii) shows that total prior uncertainty bounds the joint contribution of solicitation depth and assortment breadth. It also yields a decreasing upper bound on $\textup{VOA}(k; m)$ as $m$ grows, since solicitation reduces the remaining posterior variance. Section~\ref{sec:partition-tree} interprets this budget geometrically.

Proposition~\ref{prop:general-structure}(iii) provides a conservative benchmark for any fixed non-adaptive query sequence. For that same sequence, the Gaussian prior $\mathcal{N}(\bmu_0,\bSigma_0)$ yields the largest expected residual uncertainty. Equivalently, it yields the smallest solicitation gain, because the Gaussian posterior mean coincides with the LMMSE estimator and the Bayes MMSE is bounded above by the LMMSE. Accordingly, the Gaussian formulas provide conservative guarantees for any prior distribution satisfying Assumption~\ref{as:second-moment}, as long as the agent commits to the same non-adaptive query directions.

\subsection{The Two-Phase Partition Tree}\label{sec:partition-tree}

\begin{figure}[htp]
\centering
\includegraphics[width=\textwidth]{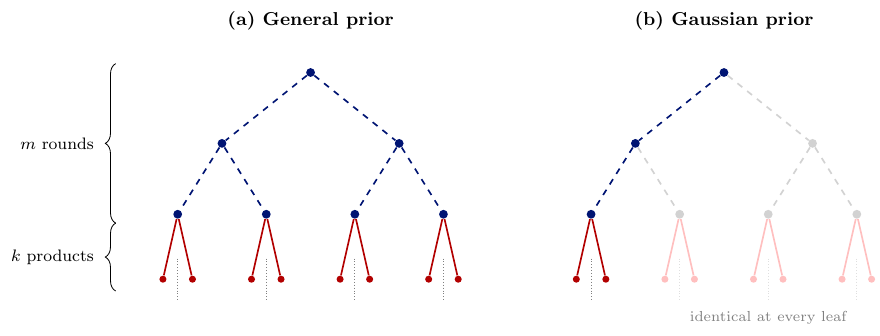}
\caption{The two-phase partition tree ($m = 2$, $k = 2$).}
\label{fig:tree-structure}
{\begin{flushleft}\footnotesize{\it Note.} Blue dashed edges are solicitation (soft, noisy), red solid edges are assortment (hard, noiseless). (a)~Under a general prior, the solicitation tree branches adaptively: different customer responses yield different posteriors, and prompt different follow-up queries. At each leaf, $k = 2$ products induce a noiseless Voronoi partition (dotted lines). (b)~Under a Gaussian prior, the posterior covariance $\bSigma_t$ is deterministic, so an optimal query can be chosen identically at every branch of a given depth. The solicitation tree can be represented by a single highlighted path, while the assortment structure is identical, up to translation by the posterior mean, at every leaf.\end{flushleft}}
\end{figure}

\begin{figure}[htp]
\centering
\includegraphics[width=\textwidth]{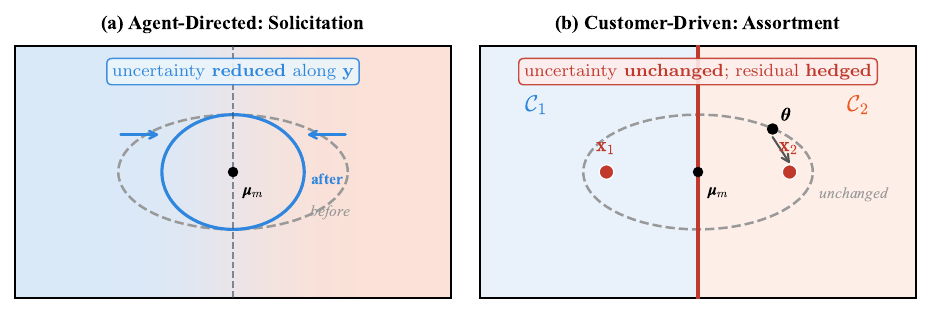}
\caption{Learning versus hedging in the two-phase partition ($d = 2$).}
\label{fig:partition-tree}
{\begin{flushleft}\footnotesize{\it Note.} (a)~Agent-directed solicitation performs a \emph{soft} hyperplane cut: the noisy response $z = \btheta^\top\by + \epsilon$ gives a probabilistic signal about which side of the hyperplane the customer falls onto, with ambiguity near the boundary controlled by the noise $\sigma$. The posterior (solid blue) \emph{shrinks} relative to the prior (dashed gray), with residual uncertainty decaying on the order $1/m$ as the solicitation budget $m$ grows (Proposition~\ref{prop:substitutes}). (b)~Customer-driven assortment performs a \emph{noiseless} Voronoi cut: the agent places $k = 2$ products (red dots), and the customer deterministically selects the nearest one. The posterior is \emph{unchanged}, before the choice is observed, but the assortment hedges the residual uncertainty. The cost asymmetry between the two phases is governed by Proposition~\ref{prop:substitutes}.\end{flushleft}}
\end{figure}

To build intuition for these structural results, we develop a geometric interpretation of the solicit-then-suggest model as a pair of nested partition trees over the preference space $\R^d$. This interpretation applies under any prior, though the Gaussian case reveals a particularly clean structure that the general case enriches. The tree viewpoint clarifies \emph{why} solicitation depth and assortment breadth are substitutes and \emph{how} preference uncertainty splits between learning via solicitation and hedging via assortment. Figure~\ref{fig:tree-structure} illustrates this structure, and Figure~\ref{fig:partition-tree} contrasts the two mechanisms spatially.

The model proceeds in two phases: a solicitation phase in which the agent queries the customer, followed by an assortment phase in which the customer selects from the recommended products. Each phase generates a partition tree, and the two compose into a single hierarchy. To illustrate, consider an agent that asks $m = 2$ questions and then offers $k = 2$ products. The solicitation phase produces a tree of depth two: the first question partitions the population by their responses, the second refines each group, yielding distinct belief states at each leaf. At each belief state, the agent places two products on opposite sides of the posterior mean to cover the remaining uncertainty, and the customer selects whichever product is closer to their ideal point.

{\bf Phase 1: the solicitation tree (agent-directed partitioning).}
The $m$-round solicitation policy generates a tree $\mathcal{T}_S$ of depth $m$. At each internal node at depth $t$, the agent holds a posterior $P_t$ and selects a Bellman-optimal query direction $\by_{t+1}^*(P_t)$. The customer's response determines which branch is taken, yielding the updated posterior $P_{t+1}$ at the child node. Each query performs a soft hyperplane cut: the observation $z_t = \btheta^\top \by_t + \epsilon_t$ localizes the customer relative to the hyperplane $\{\by_t^\top \btheta = z_t\}$, with lower noise ($\sigma^2$ small) giving a more informative response.

Under the Gaussian prior, the solicitation tree reduces to a single path: the posterior covariance $\bSigma_t$ is deterministic given the query directions (Lemma~\ref{lem:posterior}), so every branch at a given depth carries the same posterior covariance and admits the same set of optimal next queries. Under a non-Gaussian prior, the tree is branching. Different customer responses yield distinct posterior shapes, which in turn call for different follow-up questions. The query at round $t+1$ depends on the full history of responses $(z_1, \ldots, z_t)$, which makes solicitation inherently adaptive.

{\bf Phase 2: the assortment tree (customer-driven partitioning).}
Once solicitation concludes, the agent deploys a $k$-product assortment at each leaf of the solicitation tree. The $k$ products partition the remaining preference space into Voronoi cells, one per product, and the customer selects the product that is closest to their ideal point. This selection can be viewed as a one-level tree that splits the preference space into $k$ regions. Unlike the solicitation phase, each split in the assortment tree is noiseless: the customer deterministically selects the nearest product. The assortment does not reduce the agent's uncertainty before the choice is observed, but hedges it by ensuring that the selected product lies in the correct Voronoi cell. The two phases have fundamentally different cost structures. In the solicitation phase, each additional question yields a predictable reduction, with residual uncertainty scaling as $O(1/m)$ (Proposition~\ref{prop:substitutes}(i)). In the assortment phase, adding more products refines the Voronoi partition, but to cover all $d$ dimensions of residual uncertainty requires a number of products that grows exponentially in $d$. This cost asymmetry reinterprets the substitutability result of Proposition~\ref{prop:substitutes} through the partition tree: expanding the solicitation phase is polynomially efficient in the number of questions $m$, while expanding the assortment is exponentially costly in the dimensionality $d$.

\subsection{Asymptotic Gaussianity and Universality}\label{sec:gaussian-approx}

Proposition~\ref{prop:general-structure} establishes that the value decomposition and substitutability logic extend to arbitrary priors, yet the closed-form solutions in Section~\ref{sec:analysis} relied on a Gaussian prior. Corollary~\ref{cor:bvm} justifies the Gaussian benchmark more broadly: under regularity conditions, when the solicitation policy explores all preference dimensions sufficiently, the posterior converges to a Gaussian as the number of rounds grows.

\begin{corollary}[{\sc Gaussian Convergence}]\label{cor:bvm}
Fix a dimensionality $d$ and a true parameter value $\btheta^\star\in\R^d$. Consider any solicitation policy $\pi=(\pi_t)_{t\ge 1}$ that selects unit-norm queries adapted to the history,
$$
\by_t = \pi_t(\cH_{t-1}), \qquad \|\by_t\|_2=1,
$$
where $\pi_t$ may be randomized conditional on $\cH_{t-1}$. Let $P_m = P(\btheta \mid \cH_m)$ denote the posterior after $m$ rounds. Suppose the prior $P_0$ satisfies Assumption~\ref{as:second-moment} and admits a density $p_0$ that is bounded, continuous, and strictly positive at $\btheta^\star$. Define
$$
S_m := \sum_{t=1}^{m} \by_t \by_t^\top, \qquad \tilde\bSigma_m := \left(\sigma^{-2}S_m\right)^{-1}.
$$
If the realized query sequence satisfies $\lambda_{\min}(S_m)\to\infty$ and $\log m = o\!\left(\lambda_{\min}(S_m)\right)$, then in probability,
$$
d_{TV}\!\left(P_m,\, \mathcal{N}\!\left(\bthetahat_m,\, \tilde\bSigma_m\right)\right) \to 0 \quad \text{as } m \to \infty,
$$
where $\bthetahat_m = \E[\btheta \mid \cH_m]$ is the posterior mean. Under a Gaussian prior, the exact posterior covariance $\bSigma_m = \bigl(\bSigma_0^{-1} + \sigma^{-2}S_m\bigr)^{-1}$ is asymptotically equivalent to $\tilde\bSigma_m$ when $\lambda_{\min}(S_m)\to\infty$.
\end{corollary}

The result follows from the adaptive-design Bernstein-von Mises theorem of \citet{DuNairJanson2025}. The key condition is that $\lambda_{\min}(S_m)\to\infty$ with $\log m = o(\lambda_{\min}(S_m))$, which requires that even the least-informed direction in the preference space receive sufficient exploration relative to the total number of rounds. A practical way to satisfy this is to mix targeted, history-dependent queries with occasional exploratory queries drawn from a fixed spanning set. Under any such policy that explores all preference dimensions sufficiently, the posterior is asymptotically Gaussian with covariance $\tilde\bSigma_m$, so the closed-form expressions in Section~\ref{sec:analysis} provide asymptotically accurate approximations for arbitrary regular priors.

Corollary~\ref{cor:bvm} ensures that the posterior becomes Gaussian, but the covariance $\tilde\bSigma_m$ is shaped by the cumulative query allocation through $S_m$: directions that received more solicitation effort have smaller residual uncertainty. Only policies that balance information across all directions, such as the $k=1$ water-filling benchmark once every preference dimension has been queried, additionally drive $\tilde\bSigma_m$ toward isotropy. The Gaussian model, therefore, plays two roles: it yields tractable closed forms, and it serves as an asymptotically valid approximation for any regular prior under solicitation that explores all preference dimensions.

\section{Conclusion}\label{sec:conclusion}

Agentic purchasing is reshaping how customers buy products. This paper develops the solicit-then-suggest framework to formalize and analyze this emerging problem. Our study has several limitations that provide opportunities for future research.

First, our model takes the observation technology as given: the agent chooses which direction to query but not \emph{how} to solicit information. In practice, the agent also selects among question formats, such as pairwise comparisons, rating scales, and visual previews, each of which induces a different noise structure and information content per round. Joint optimization of the format and content of solicitation introduces a richer information design problem in which the noise variance $\sigma^2$ itself becomes endogenous and may depend on the question type and the customer's cognitive constraints. A central question is how the agent should co-design the solicitation format and query direction to maximize preference learning. This question connects to the growing literature on information design and the practical challenge of human-AI interaction design for AI shopping assistants.

Second, this paper takes a first step toward modeling agentic purchasing by focusing on maximizing customer match quality. In practice, each product may carry a different commission or margin for the agent, creating a tradeoff between recommending the best fit for the customer and recommending products with higher rewards for the agent. This reward-aware assortment problem relates to classical assortment optimization, in which a firm balances variety against profitability, but now the agent has an additional instrument, solicitation, that shapes the information available at the time of the assortment decision. Beyond rewards, sellers may strategically set or adjust prices in response to the agent's behavior, making pricing an endogenous part of the purchasing process and raising new questions about mechanism design when an AI agent intermediates between customers and sellers. This relates to the active literature on algorithmic pricing and platform design.

\bibliographystyle{ormsv080}
\bibliography{main}

\end{document}